\documentclass[pra,aps,twocolumn,amsmath,footinbib]{revtex4-2}
\usepackage{amssymb}
\usepackage{graphicx}
\usepackage{hyperref}

\usepackage{color} 

\definecolor{darkgreen}{rgb}{0,0.7,0}  

\usepackage[normalem]{ulem} 

\usepackage[table]{xcolor}
\definecolor{lightgray}{gray}{0.9}

\usepackage{booktabs}

\usepackage[caption=false]{subfig}
\newcommand{\phantomsubfloat}[1]{
    {
        \captionsetup[subfigure]{labelformat=empty}
        \subfloat[][]{#1}
    }%
}

\begin{document}


\title{Characterising transport in a quantum gas by measuring Drude weights}


\author{Philipp Sch{\"u}ttelkopf}\email{philipp.schuettelkopf@tuwien.ac.at}
\author{Mohammadamin Tajik}
\author{Nataliia Bazhan}
\author{Federica Cataldini}
\author{Si-Cong Ji}
\author{J\"org Schmiedmayer}
\author{Frederik M{\o}ller}\email{frederik.moller@tuwien.ac.at}
\affiliation{
Vienna Center for Quantum Science and Technology (VCQ), Atominstitut, TU Wien, Vienna, Austria}

\date{\today}

\begin{abstract} 
Transport properties play a crucial role in defining materials as insulators, metals, or superconductors.
A fundamental parameter in this regard is the Drude weight, which quantify the ballistic transport of charge carriers.
In this work, we measure the Drude weights of an ultracold gas of interacting bosonic atoms confined to one dimension, characterising the induced atomic and energy currents in response to perturbations with an external potential.
We induce currents through two distinct experimental protocols; by applying a constant force to the gas, and by joining two subsystems prepared in different equilibrium states. 
By virtue of integrability, dynamics of the system is governed by ballistically propagating, long-lived quasi-particle excitations, whereby Drude weights almost fully characterise large-scale transport. 
Indeed, our results align with predictions from a recently developed hydrodynamic theory, demonstrating almost fully dissipationless transport, even at finite temperatures and interactions.
These findings not only provide experimental validation of the hydrodynamic predictions but also offer methodologies applicable to various condensed matter systems, facilitating further studies on the transport properties of strongly correlated quantum matter.
\end{abstract} 

\maketitle

Drude weights are fundamental transport coefficients of condensed matter physics, often used for classifying materials according to their conductivity properties~\cite{PhysRevB.47.7995}.
Initially formulated for solid state systems, where it is also referred to as charge stiffness, the Drude weight measures the ratio of the density of mobile charge carriers to their mass~\cite{Drude_Metalle}. 
In the context of quantum many-body systems, Drude weights quantify the ballistic transport of charge carriers and, for non-vanishing values, signal the presence of persistent dynamical correlations.
Hence, the scaling behavior of Drude weights in the thermodynamic limit is an essential criterion for distinguishing metallic and insulating phases in many-body systems; in the absence of impurities, metals exhibit a finite Drude weight, whereas it vanishes for insulators~\cite{PhysRev.133.A171}.

At finite temperatures, Drude weights of interacting systems are generally expected to vanish, however, there are notable exceptions with anomalous transport properties.
Among the most studied examples are graphene~\cite{PhysRevB.83.165113, PhysRevB.84.045429, PhysRevLett.113.056602}, whose linear dispersion allows thermally excited carriers to contribute to conductivity, and integrable models~\cite{PhysRevB.53.983, PhysRevLett.74.972}, where a macroscopic number of conservation laws prevents the complete decay of currents.
Recently, the development of a hydrodynamic theory for integrable systems, dubbed Generalized Hydrodynamics (GHD)~\cite{PhysRevX.6.041065, PhysRevLett.117.207201}, has advanced the study of transport and dynamics in several paradigmatic models of quantum many-body physics~\cite{PhysRevB.97.045407, De_Nardis_2022, doyon2023generalized}. 
In these models, conserved charges are carried by interacting quasi-particles representing collective excitations with infinite lifetimes, even at finite temperatures, which propagate mainly ballistically with sub-leading diffusive corrections~\cite{PhysRevLett.121.160603, PhysRevB.98.220303}.
Therefore, Drude weights may fully characterise the large-scale dynamics of integrable systems.
However, despite theoretical progress, direct measurements of transport coefficients have yet to be performed, even though several experiments have shown hydrodynamic evolution consistent with the theoretical predictions~\cite{schemmer2019generalized, malvania2020generalized, PhysRevLett.126.090602, PhysRevX.12.041032}.

\begin{figure*}
    \centering
    \includegraphics{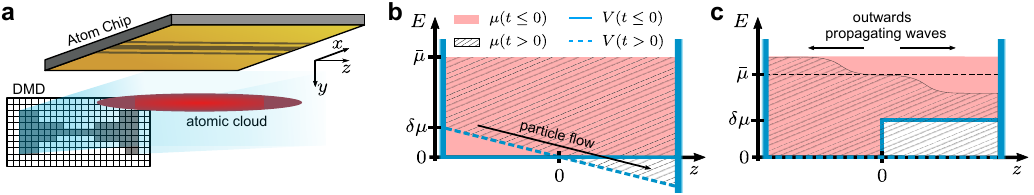}
    
    \phantomsubfloat{\label{fig:illustration_a}}
    \phantomsubfloat{\label{fig:illustration_b}}
    \phantomsubfloat{\label{fig:illustration_c}}
    \vspace{-2\baselineskip}

    \caption{\label{fig:illustration}
    \textbf{Schematic of the experimental protocol.}
    \textbf{a,} An Atom Chip creates a tight transverse confinement (along the $x$ and $y$ axes), confining a gas of ultracold bosons to one dimension. Along the longitudinal $z$-axis, arbitrary potentials can be realised using a Digital Micromirror Device (DMD).
    \textbf{b,} An initially homogeneous system with chemical potential $\bar\mu = g_{1\mathrm{D}} \bar n$ is realised by confining the 1D gas in a box trap. At time $t = 0$ the bottom of the box trap $V(z)$ is tilted, leading to an accelerating current of atoms across the center $z = 0$.
    \textbf{c,} An initial chemical potential imbalance $\delta\mu$ is achieved by creating a box trap with a step-function bottom. At time $t = 0$ the trap bottom is quenched to flat, leading to two counter-propagating waves originating from the center.   
    }
\end{figure*}

In this work, we measure Drude weights of an ultracold gas of interacting bosonic atoms confined to one dimension, making the system integrable.
Analogous to their solid state counterparts, the Drude weights of the gas are extracted by measuring the current of particles and energy in response to controlled perturbations.
We accomplish this through the implementation of two distinct protocols: First, probing the response of the gas to a constant, external force and, second, measuring the flow of charges between two subsystems in different equilibrium states.
In accordance with theoretical predictions, we find that the Drude weights almost entirely characterise the observed dynamics of the gas, despite its finite temperature and interaction dominated nature.
Further, the methodologies that we have employed can be generalised to a number of condensed matter systems in a straightforward manner, thus offering a means to probing the emergent transport behavior of strongly correlated quantum systems.

Our experimental system is a one-dimensional (1D) ultracold Bose gas trapped below an Atom Chip~\cite{reichel2011atom} (Fig.~\ref{fig:illustration_a}).
The chip produces a cigar-shaped magnetic trap with a trapping frequency of $\omega_\perp = 2 \pi \times 1.38 \, \mathrm{kHz}$ along the two transverse directions, strong enough to confine the atoms to a single dimension.
We prepare a thermal equilibrium state by cooling an atomic cloud of $^{87}$Rb atoms with mass $m$ using standard techniques of laser and evaporative cooling, resulting in temperatures typically in the range 10-50~nK.
The interaction strength is quantified by the dimensionless Lieb-Liniger parameter $\gamma = m g_{1\mathrm{D}} / \hbar^2 \bar n$~\cite{lieb1963exact}, where $g_{1\mathrm{D}}$ is the effective atomic coupling strength~\cite{PhysRevLett.81.938} and $\bar n$ is the mean atomic density. 
In our experiment, $\gamma \approx 0.002$, which, combined with the low temperatures, places the gas deeply in the quasi-condensate regime.
Hence, the energy density is dominated by interactions $\varepsilon \gtrsim \frac{1}{2} g_{1\mathrm{D}} n^2 g^{(2)}(0)$, where $g^{(2)}(0) \approx 1$ is the local two-body autocorrelation function~\cite{PhysRevLett.91.040403}, and the atomic density is approximately linearly proportional to the chemical potential $\mu = g_{1\mathrm{D}} n$.
To manipulate the gas, we utilize a Digital Micromirror Device (DMD) to generate fully customised optical dipole potentials~\cite{Tajik:19}, which are superimposed onto the magnetic trap along the longitudinal $z$-axis, see Fig.~\ref{fig:illustration_a}.
In all of our experiments, two hard walls are imposed on the condensate, confining it to a region of typical length $L = 100\,\mu\mathrm{m}$.
The system length $L$ sets the maximal duration of the experiments, as atoms reflected off the walls move against the induced current. 
Finally, dynamics are instigated by quenching the potential between the walls, denoted as $V(z)$; the subsequent evolution of the system is probed by measuring the atomic density following different evolution times $t$ using absorption imaging.

The conductivity of a system $\sigma_{ab}$ characterises the linear response of the current of the operator $b$ when coupling an external field to the operator $a$.
For a homogeneous system, its real part reads~\cite{mahan2013many} 
\begin{equation}
    \mathrm{Re} \: \sigma_{a b}(\omega) = 2 \pi \mathfrak{D}_{a b} \delta(\omega) + \sigma_{a b}^{\mathrm{reg}}(\omega) \; ,
    \label{eq:conductivity}
\end{equation}
where $\sigma_{a b}^{\mathrm{reg}} (\omega)$ denotes the regular frequency-dependent part, while the singular part indicates a ballistic (dissipationless) contribution, whose magnitude is specified by the Drude weight $\mathfrak{D}_{a b}$.
From the perspective of hydrodynamics, the regular or dissipative part of the conductivity encodes diffusive processes, which in integrable systems arise from thermal fluctuations~\cite{PhysRevLett.121.160603, PhysRevB.98.220303} or weak breaking of integrability~\cite{PhysRevB.101.180302}.
The former can generally be excluded for quasi-condensates~\cite{10.21468/SciPostPhysCore.7.2.025}, as diffusion is irrelevant on experimental timescales (Methods).
Further, previous studies of our system show that mechanisms of integrability breaking are weak on experimental time scales~\cite{PhysRevLett.126.090602, PhysRevX.12.041032}.
Hence, Drude weights fully characterise the transport of the condensate.
The conventional way of expressing $\mathfrak{D}_{a b}$ is via the Kubo formula~\cite{Kubo} using the time-averaged current-current autocorrelation function (Methods).
However, this definition is inapproachable from an experimental standpoint; the destructive nature of absorption imaging makes measuring dynamical autocorrelation functions highly challenging, thus necessitating protocols where the Drude weight can be accessed otherwise.

\begin{figure}
    \centering
    \includegraphics[width = 1\columnwidth]{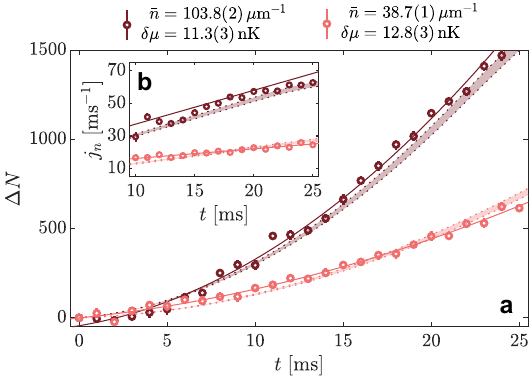}

    \phantomsubfloat{\label{fig:imbalance_a}}
    \phantomsubfloat{\label{fig:imbalance_b}}
    \vspace{-2\baselineskip}

    \caption{\label{fig:imbalance}
    \textbf{Particle flow in response to constant external force.}
    \textbf{a,} Measured difference of atoms between the left and right halves of the system for two experiments with different mean densities and potential gradients, where $\delta\mu = \phi_n L/2$. Quadratic fits of the imbalances are plotted as solid lines, while shaded areas indicate results of GHD simulations accounting for the uncertainty in $\bar n$ and $\delta\mu$. 
    \textbf{b,} Atomic currents obtained from the time derivative of the fitting functions. 
    The measured current is approximated as $\Delta N/t$.
    Error bars indicate the standard error of the mean and numbers in brackets represent statistical uncertainties over 40 repititions.
    }
\end{figure}

A key signature of the Drude weight is a dissipationless current, whereby it may alternatively be defined as the time-asymptotic growth rate of the \textit{local} current in response to a small, time-independent external field~\cite{Ilievski2013} 
\begin{equation}
    \mathfrak{D}_{ab} = \lim_{\phi_a \to 0} \lim_{t \to \infty} \frac{ j_b (z, t) }{t \phi_a} \; ,
    \label{eq:Drudeweight_constforce}
\end{equation}
where $\phi_a$ is the gradient of the field coupling to the operator $a$.
In our first experiment, we implement such a protocol starting from a flat potential between the walls, thus preparing a homogeneous gas; at time $t = 0$ the flat potential is switched to a constant gradient $V(z) = \phi_n z$, equivalent to creating a linear electric field across a charge carrier (Fig.~\ref{fig:illustration_b}).
We extract the local atomic current in the center of the system by measuring the imbalance of atoms between the left and right side of the system $\Delta N (t) = N_L (t) - N_R (t)$, such that $ j_n (z = 0, t) = \frac{1}{2} \frac{\mathrm{d}  \Delta N(t) }{\mathrm{d} t}$, repeating the experiment for various mean densities $\bar n$ and potential gradients $\phi_n$.

\begin{figure*}
    \centering
    \includegraphics[width = 1\textwidth]{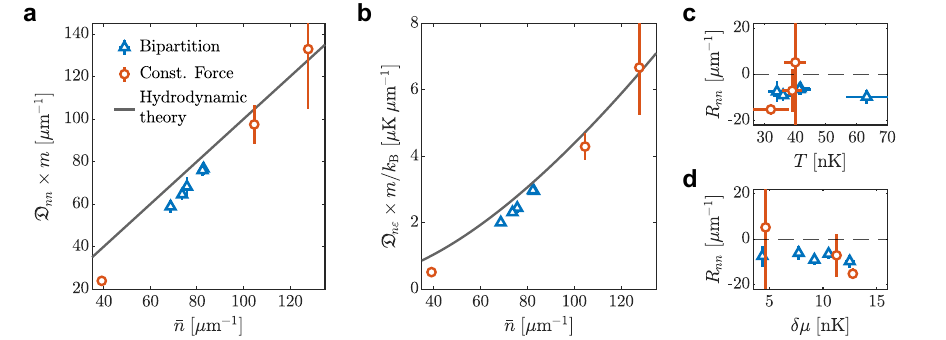}

    \phantomsubfloat{\label{fig:Drude_weights_a}}
    \phantomsubfloat{\label{fig:Drude_weights_b}}
    \phantomsubfloat{\label{fig:Drude_weights_c}}
    \vspace{-2\baselineskip}

    \caption{\label{fig:Drude_weights}
    \textbf{Comparison of Measured Drude Weights with hydrodynamic theory.}
    \textbf{a,b,} Drude weights characterising the linear response of the atomic current \textbf{(a)} and energy current \textbf{(b)} to a field coupled to the atomic density. The experimentally measured coefficients from the constant force and bipartition protocols are derived from Eqs.~\eqref{eq:Drudeweight_constforce} and~\eqref{eq:Drudeweight_bipartition}, respectively, while the theoretical predictions for a thermal states at $T = 50\:\mathrm{nK}$ are calculated from Eq.~\eqref{eq:Drudeweight_GHD}. Results are plotted as a function of the background atomic density $\bar{n}$.
    \textbf{c,d,} Residuals of the measurements compared to the GHD theory are shown as a function of temperature $T$ \textbf{(c)} and perturbation $\delta\mu$ \textbf{(d)}, where $\delta\mu = \phi_n L/2$ for the constant force measurements.
    Error bars indicate standard error.
    }
\end{figure*}

Fig.~\ref{fig:imbalance_a} shows the measured imbalance dynamics of two such experiments; as expected, larger densities and gradients lead to a more rapid growth of $\Delta N$.
To extract the current, we fit the measured imbalances with a second order polynomial.Examples of the imbalance fits and their derivatives are shown in Figs.~\ref{fig:imbalance_a} and~\ref{fig:imbalance_b}, respectively.
Given a linear growth of the current, it may alternatively be expressed as $j_n = \Delta N/t$, which, for the measured imbalance, is plotted in Fig.~\ref{fig:imbalance_b} and aligns with the fit, particularly at late times.
At short timescales, the measured current is somewhat irregular owing to the initial density profile not being completely homogeneous (Methods).
Since the energy of the condensate is dominated by interactions, the local energy current can be expressed in terms of the atomic current as $j_\varepsilon = g_{1\mathrm{D}} n j_n$, thus enabling us to calculate the corresponding Drude weight via Eq.~\eqref{eq:Drudeweight_constforce}.
The results, plotted in Fig.~\ref{fig:Drude_weights}, will be discussed after explaining the second protocol. 

\begin{figure}
    \centering
    \includegraphics[width = 1\columnwidth]{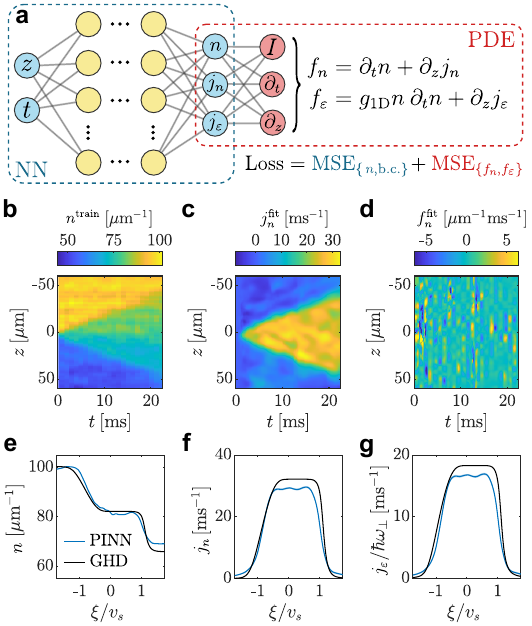}

    \phantomsubfloat{\label{fig:PINN_fitting_a}}
    \phantomsubfloat{\label{fig:PINN_fitting_b}}
    \phantomsubfloat{\label{fig:PINN_fitting_c}}
    \phantomsubfloat{\label{fig:PINN_fitting_d}}
    \phantomsubfloat{\label{fig:PINN_fitting_e}}
    \phantomsubfloat{\label{fig:PINN_fitting_f}}
    \phantomsubfloat{\label{fig:PINN_fitting_g}}
    \vspace{-2\baselineskip}

    \caption{\label{fig:PINN_fitting}
    \textbf{Extracting bipartition currents with Physics Informed Neural Networks (PINNs).}
    \textbf{a,} The PINN has two components: \textit{(i)} A neural network that outputs local atomic density $n$, atomic current $j_n$, and energy current $j_\varepsilon$ given input $z, t$, and \textit{(ii)} Partial differential equations for mass $f_n$ and energy $f_\varepsilon$ continuity. By minimizing the mean square error (MSE) of the continuity equations, boundary conditions, and the difference between measured and output density, the currents are inferred.
    \textbf{b,c,d,} Measured mean atomic density over 65 repetitions used for training \textbf{(b)}, along with predicted atomic current \textbf{(c)} and atomic continuity equation \textbf{(d)} obtained from averaging the output of ten independently trained networks.
    \textbf{e,f,g,} At large scale, the atomic density profile \textbf{(e)}, atomic current profile \textbf{(f)}, and energy current profile \textbf{(g)} reach a self-similar solution along $\xi = z/t$. The PINN fits, averaged over intermediate times $t$ and ten models, are compared to GHD simulations.
    }
\end{figure}

Our second experiment implements the bipartition protocol, in which two subsystems, initially in equilibrium with a chemical potential imbalance $\delta \mu$, are joined at a point contact, inducing particle flow between them. 
Unlike the flow of charges through a 1D channel connecting two imbalanced reservoirs, which has been studied in experiments with ultracold Fermions~\cite{doi:10.1126/science.1223175, doi:10.1126/science.aac9584}, the dynamics of the bipartiton protocol is determined solely by the bulk transport properties of the system.
The extraction of Drude weights from the ensuing dynamics is facilitated by the recently developed theory of Generalized Hydrodynamics~\cite{PhysRevX.6.041065, PhysRevLett.117.207201}, whose inception has heralded significant progress in understanding one-dimensional transport (Methods).
Following hydrodynamic evolution in the bipartition, a local quasi-stationary state emerges along each ray $\xi = z/t$; the Drude weight is given in terms of the \textit{total} asymptotic current via~\cite{PhysRevLett.119.020602}
\begin{equation}
    \mathfrak{D}_{nb} = \lim_{\delta\mu \to 0} \frac{1}{ 2 \: \delta\mu} \int \mathrm{d}\xi \:   j_b (\xi)   \; .
    \label{eq:Drudeweight_bipartition}
\end{equation}
To realise the bipartition protocol, we first prepare a thermal state in the potential $V(z) = \delta\mu \, \Theta (z)$, where $\Theta (z)$ is the Heaviside step function, such that the left and right half sides of gas have different densities.
At time $t=0$~ms we quench to a flat potential and observe in the ensuing dynamics two outwards counter-propagating waves (Fig.~\ref{fig:illustration_c}).

In order to obtain both the atomic and energy currents from the sparsely measured atomic density, we employ a Physics Informed Neural Network (PINN)~\cite{RAISSI2019686}.
Briefly, given sparse or noisy measurements, PINNs leverage knowledge of the governing physical equations, here continuity of mass and energy, to accurately infer the underlying fields.
By ensuring consistency with the conservation laws, an enhanced fidelity of the reconstructed currents compared to methods based on finite difference can be achieved.  
See Methods for further details.
An illustration of our PINN model is depicted in Fig.~\ref{fig:PINN_fitting_a}; given an input set of coordinates $(z, t)$, the model predicts the local atomic density $n$, atomic current $j_n$, and energy current $j_\varepsilon$.
Figs.~\ref{fig:PINN_fitting}(b-d) show the measured density of one bipartition experiment used for training, the corresponding atomic current predicted by the model, and the mass continuity of the predictions, respectively. 
Despite lacking direct knowledge of the currents beyond their boundary conditions, the model infers them with high accuracy, as evidenced by adherence to the conservation laws and agreement with GHD simulations~\cite{10.21468/SciPostPhys.8.3.041, MOLLER2023112431} of the dynamics (Methods) shown in Figs.~\ref{fig:PINN_fitting}(e-g).
Across multiple realisations of the bipartition protocol, a slight discrepancy in the amplitude of the quasi-stationary currents along the ray $\xi = z/t$ consistently emerges between the GHD and PINN predictions. 
We attribute this deviation to imperfections in the optical potential, which introduce density variations in the initial state and disorder in the flat potential~\cite{doi:10.1126/science.1223175}; both not accounted for in the GHD simulations. 

Finally, having extracted the atomic and energy currents from both the constant force and bipartition protocols, we compute the corresponding Drude weights via Eqs.~\eqref{eq:Drudeweight_constforce} and ~\eqref{eq:Drudeweight_bipartition}, respectively.
Formally, these expressions are exact only in the limits of asymptotic time and infinitesimal perturbations.
The former limit ensures that only dissipationless currents are accounted for.
However, due to the negligible diffusion in the quasi-condensate, transport at accessible time scales is entirely ballistic.
Meanwhile, the latter limit reflects the linear response of the conductivity; although the experimental perturbations are finite, GHD simulations confirm that the responses remain within the linear regime such that Eqs.~\eqref{eq:Drudeweight_constforce} and ~\eqref{eq:Drudeweight_bipartition} accurately yield the Drude weights under experimental conditions (see Appendix). 

The Drude weights obtained from experimental measurements are presented in Figs.~\ref{fig:Drude_weights}(a,b) as a function of mean density $\bar n$, alongside theoretically predicted values for homogeneous thermal states at 50~nK, derived from the Kubo formalism using hydrodynamic projections~\cite{10.21468/SciPostPhys.3.6.039} (Methods).  
Analogously to solid state systems, where the Drude weight measures the density of mobile charge carriers relative to their mass, we find that $\mathfrak{D}_{n n}$, which characterizes the atomic current response to a density perturbation, is proportional to $\bar n/m$, consistent with theoretical predictions.
This surprisingly simple scaling behavior, particularly in an interacting system at finite temperature, arises from a uniform and interaction-independent coupling of the external potential to the quasi-particle excitations of the gas~\cite{10.21468/SciPostPhys.2.2.014}. 
Moreover, $\mathfrak{D}_{n \varepsilon}$, which quantifies the ballistic energy current induced by coupling to the atomic density, also shows agreement with theory, notwithstanding that the theory accounts for the total energy, whereas only the interaction energy current is measured. 
At lower atomic densities, the interaction energy scales quadratically with $\bar n$ density, while at higher densities, the effective 1D coupling $g_{1\mathrm{D}}$ acquires a weak density dependence~\cite{PhysRevA.65.043614}, leading to the energy becoming more linear in its dependence on $\bar n$.

For both protocols studied, the experimentally extracted Drude weights are generally slightly lower than theoretical predictions.
Plotting the residuals $R_{n n} = \left(\mathfrak{D}_{n n}^{\mathrm{meas}} - \mathfrak{D}_{n n}^{\mathrm{theory}} \right) m$ as a function of temperature and perturbation amplitude reveal no dependence on either parameter (see Figs.~\ref{fig:Drude_weights}(c,d)).
Since the GHD prediction is calculated for a fixed temperature of 50~nK, the former demonstrates that, within quasi-condensate regime, the probed responses of the gas are independent of temperature.
Meanwhile, the latter confirming the linear response of the system for the given perturbations.
Thus, we attribute the observed discrepancies in the measured Drude weights to imperfections in the optical dipole potential, which effectively reduce the conductivity of the system.

In summary, we have measured the Drude weights of a 1D quasi-condensate, quantifying the ballistic transport of mass and energy when coupling to the atomic density via an external potential.
The simplicity of the 1D Bose gas makes it an ideal platform for verifying GHD predictions of transport coefficients~\cite{10.21468/SciPostPhys.3.6.039, PhysRevLett.119.020602}, however, our methodologies can be easily adapted to other systems, providing a foundation for future studies of quantum matter transport properties.
Integrable models such as the XXZ spin-1/2 chain and the sine-Gordon model exhibit anomalous transport~\cite{Bulchandani_2021} with Drude weights showing fractal-like dependence on interactions~\cite{PhysRevLett.106.217206, PhysRevB.108.L241105}.
Additionally, GHD extends to non-thermal equilibrium states supported by integrable models~\cite{PhysRevLett.98.050405, doi:10.1126/science.1257026}, whose transport properties remain experimentally unexplored.
Finally, introducing non-integrable perturbations has shown to significantly alter long-term dynamics, transitioning from integrable to conventional hydrodynamics and resulting in a dynamical phase transition in transport properties~\cite{PhysRevX.8.021030, PhysRevLett.126.090602, Bastianello_2021}.

\subsection*{Acknowledgements}
We thank Sebastian Erne and Igor Mazets for helpful discussions. 
This research was supported by the European Research Council: ERC-AdG ``Emergence in Quantum Physics'' (EmQ) under Grant Agreement No. 101097858 and the Austrian Science Fund (FWF) ``Strongly correlated Quantum Fields out of equilibrium'' [grant DOI: 10.55776/P36656].

\appendix

\section{Experimental protocol}

Using standard techniques of laser cooling, magnetic trapping, and evaporative cooling we load $\mathrm{Rb}^{87}$ atoms in a tight transversal confinement with trapping frequency $\omega_\perp = 2 \pi \times 1.38 \, \mathrm{kHz}$ with $\mu, k_BT < \hbar \omega_\perp$ on an Atom Chip~\cite{reichel2011atom}.
The initial thermal states are realised by evaporatively cooling the atoms while trapped in the combined magnetic and optical dipole potential. 
By tuning the evaporative cooling, we can realise a gas with number of atoms $N$ in the range $(3-10)\times 10^3$ at temperatures (10-50) nK, the latter measurable via density-ripples thermometry~\cite{PhysRevA.81.031610}. Typical shot to shot density variation after post selection is $7-10 \%$, while heating and atom losses ($\sim0.8\%$) are negligible over the time scales of our protocols. Nonetheless, we rescale the mean density of every measure evolution time $t>0$ to equal that at $t = 0$.

Using a DMD we can create an arbitrary blue detuned optical dipole potential along the longitudinal z-axis of the trap~\cite{Tajik:19}.
To this end, we employ an optimization algorithm, which toggles individual DMD pixels on or off until a desired profile is reached. 
Due to technical limitations of this approach~\cite{Tajik:19}, the optimized density profiles feature systematic imperfections from the desired profile with typical standard deviation $\sigma_n = 0.05 \bar n$.
Further, because the optical potential needs to compensate the underlying harmonic confinement created by the chip, the length of the boxes created is limited by our available laser power.
To instigate dynamics, we rapidly quench (on time scales $\sim 20 \mu $s) between pre-optimised patterns, then hold the condensate for a desired evolution time.
Finally, the 1D density profiles are measured using absorption imaging after 2 ms free expansion time; we repeat the experiment for different evolution times and average the measurements over multiple realizations to obtain $n(z,t)$ and $\bar{n}$.
The gradients of the constant force fields $\phi_n$, the initial chemical potential imbalances $\delta \mu$ of the bipartition, and the condensate temperature $T$ are all extracted from separate measurements, taken of thermal equilibrium states in the corresponding potential configurations.

\section{Lieb-Liniger model}

Following confinement to one dimension, a gas of $N$ bosons with mass $m$ and contact interactions is well described by the integrable Lieb-Liniger Hamiltonian~\cite{lieb1963exact} plus an additional external longitudinal potential $V(z)$  
\begin{equation}
    \mathcal{H}=-\frac{\hbar^{2}}{2 m} \sum_{i=1}^{N} \partial_{z_{i}}^{2}+g_{1\mathrm{D}} \sum_{i<j} \delta\left(z_{i}-z_{j}\right)+\sum_{i=1}^{N} V\left(z_{i}\right) \; .
\end{equation}
The 1D contact interaction is depends on the transverse trapping confinement via $g_{1\mathrm{D}} = 2 \hbar \omega_{\perp} a_s \varpi \left(1-1.03 a_s/l_{\perp} \right)^{-1}$ ~\cite{PhysRevLett.81.938}, with $a_s$ being the 3D scattering length and $l_{\perp} = \sqrt{\hbar /(m \omega_\perp)}$.
The factor $\varpi =\frac{1}{2} \frac{2+3 a_s  n}{\left(1+2 a_s n\right)^{3 / 2}}$ accounts for a reduction of the 1D coupling following a transverse broadening of the bosonic wavefunction as $n a_s$ approaches unit~\cite{PhysRevA.65.043614}.
Technically, the density dependence of coupling spoils the integrability of the system.
However, we find that employing the integrable theory using an effective contact interaction evaluated for the mean density of the gas $\bar n$ accurately describes our experimental observations.
Similar observations have been made in bosonic 1D dipolar quantum gases near equilibrium, where a description using the Lieb-Lininger Hamiltonian is still feasible, with the integrability-breaking dipole interactions simply leading to an effective modification of the contact interaction~\cite{PhysRevA.107.L061302}.

\section{Extracting currents with Physics Informed Neural Networks}


The neural network of our PINN model consists of two input neurons representing the coordinates $(z,t)$ and three output neurons corresponding to the atomic density $n$, atomic current $j_n$, and energy current $j_\varepsilon$, as illustrated in Fig.~\ref{fig:PINN_fitting_a}.
The input and output layers are connected by 6 hidden layers of 30 neurons each.
Following conservation of mass and energy, the dynamics of the 1D Bose gas in our experiment is subject to the continuity equations 
\begin{equation}
\begin{aligned}    
    &\partial_t n + \partial_z j_n = 0 \; , \\
    &\partial_t \varepsilon + \partial_z j_\varepsilon = 0 \; ,
\end{aligned}
\label{eq:PINN_PDEs}
\end{equation}
where the energy density is approximated as the quasi-condensate interaction energy density $\varepsilon = \frac{1}{2} g_{1\mathrm{D}} n^2$.

The model is trained to satisfy the given training data as well as the imposed governing equations~\eqref{eq:PINN_PDEs} and boundary conditions~\cite{RAISSI2019686}.
The derivatives of the output fields with respect to input coordinates are obtained via Automatic Differentiation~\cite{JMLR:v18:17-468}.
Thus, given experimental measurements of the atomic density profile, arranged as the set $\{ z_i, t_i, n_i \}_{i = 1}^{M}$, we minimise the mean squared error (MSE) loss
\begin{equation}
\begin{aligned}
    \mathrm{Loss} =& \frac{1}{M} \sum_{i=1}^{M} \left( n(z_i, t_i) - n_i \right)^2 \\
    &+ \frac{1}{M} \sum_{i=1}^{M} \left( | f_n(z_i, t_i) |^2 + | f_\varepsilon(z_i, t_i) |^2 \right) + \mathrm{MSE}_{\mathrm{b.c.}} \; ,
\end{aligned}
\label{eq:PINN_loss}
\end{equation}
where $f_n = \partial_t n + \partial_z j_n$ and $f_\varepsilon = g_{1\mathrm{D}} n \partial_t n + \partial_z j_\varepsilon$ encode the continuity equations.
The final term of Eq.~\eqref{eq:PINN_loss} enforces boundary conditions of the currents, namely that they must vanish at the edges of the box trap, which is necessary to ensure that their solution is unique.

We find that measurement noise causes training of our model to take an excessive amount of time to converge, whereby the density profiles used for training have been convolved with a Gaussian, whose width is twice the resolution of our imaging system.
This reduces noise, resulting in a faster convergence, without affecting the observed large scale dynamics.
Further, for each data set, we independently train ten model starting from different, random parameters of the neural network; the reported currents are averaged over the different models, while errors are given by the standard error of the mean.

\begin{figure*}
    \centering
    \includegraphics{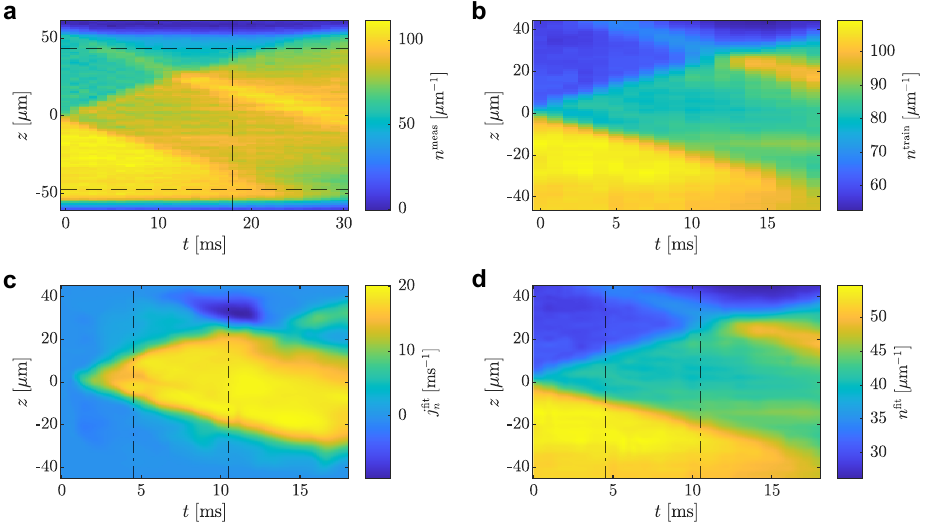}

    \phantomsubfloat{\label{fig:PINN_SM_a}}
    \phantomsubfloat{\label{fig:PINN_SM_b}}
    \phantomsubfloat{\label{fig:PINN_SM_c}}
    \phantomsubfloat{\label{fig:PINN_SM_d}}
    \vspace{-2\baselineskip}

    \caption{\label{fig:PINN_SM}
    \textbf{Example of current fitting using PINNs.}
    \textbf{a,} Measured evolution of atomic density $n$ in the bipartition protocol. The dashed lines indicate the coordinates $(z_i, t_i)$ at which $n$ is cropped to create the training data. 
    \textbf{b,} Training data for the PINN, after Gaussian smoothening.
    \textbf{c,d,} Fitted atomic current \textbf{(c)} and density\textbf{(d)}, obtained by averaging the outputs of ten PINNS trained independently on the data in \textbf{(b)}. The output is evaluated at coordinates with ten times the resolution of the training data. The dash-dotted lines indicate the time interval for which the self-similar solution along the ray $\xi = z/t$ is obtained.
    }
\end{figure*}

Having extracted the currents, their solution along the ray $\xi = z/t$ is obtained via interpolation for evolution times $t$ within an particular interval:
The self-similar solution is reasonably approached after 4~ms, whereby $\xi$ is evaluated for only later times.
Further, when quenching trap configurations from the step to the flat-bottomed box, a small perturbation is launched from the wall on the low-density side.
To exclude its contribution to the solution on the ray, only evolution times $t$ before the perturbation reaches the front of the current profile are employed (see Fig.~\ref{fig:PINN_SM}  for example).

The PINN model, which we have employed, can easily be extended to arbitrary regimes of the Bose gas (where the energy is not dominated by interactions) by including the momentum-continuity equation $\partial_t p + \partial_z j_p = 0$, where the momentum current is equal to the pressure $j_p = \mathcal{P}$~\cite{10.1063/1.1664947}.
The momentum density is given by the atomic current $p = m j_n$ following Galilean invariance of the Lieb-Liniger model, thus need not be measured.
However, knowledge of the pressure near the walls of the box trap is required to fix the boundary conditions of the PINN, as the pressure does not vanish at the boundaries of the box trap.

\section{Generalized Hydrodynamics of the 1D Bose gas}

Viewed from the perspective of the Bethe Ansatz solution, the 1D gas of neutral bosons is actually not so far from electronic solid state systems; the elementary collective excitations are fermionic quasi-particles and holes, which carry not electrical, but quanta of a macroscopically large set of conserved charges/quantities.
Each particle is uniquely labeled by its rapidity, or quasi-momentum, $\theta$, while the local distribution of quasi-particles $\rho_{\mathrm{p}} (\theta)$ fully encodes the local thermodynamic state~\cite{lieb1963exact}.
Introducing the density of allowed rapidities $\rho_{\mathrm{s}}$ and the occupations function $\vartheta (\theta) = \rho_{\mathrm{p}} (\theta) / \rho_{\mathrm{s}} (\theta)$, a thermal distribution is given by
\begin{equation}
    \vartheta (\theta) = \frac{1}{1 + e^{ \epsilon(\theta) \beta }} \; ,
\end{equation}
following the fermionic statistics of the quasi-particles.
Here, $\beta = 1/k_B T$ is the inverse temperature, and the pseudo-energy $\epsilon(\theta)$ is given by the Yang-Yang equation~\cite{10.1063/1.1664947}
\begin{equation}
    \epsilon(\theta) = \frac{\hbar^2 \theta^2}{2 m} - \mu + \frac{1}{2 \pi \beta} \int_{-\infty}^{\infty} \mathrm{d}\theta' \: \Delta\left(\theta-\theta^{\prime}\right) \ln \left( 1 + e^{ \epsilon(\theta')  \beta } \right) \; ,
\end{equation}
where $\Delta$ is the rapidity derivative of the two-body scattering phase given by
\begin{equation}
   \Delta\left(\theta-\theta^{\prime}\right) = \frac{2 m g_{1\mathrm{D}} / \hbar^2}{(m g_{1\mathrm{D}} / \hbar^2)^{2}+(\theta-\theta')^{2}} \; .
\end{equation}
In the presence of an external potential $V(z)$ one can locally shift the chemical potential accordingly $\mu (z) = \mu_0 - V(z)$ under the local density approximation, thus yielding a spatially dependent quasi-particle distribution.

Integrable systems abide to an infinite number of conservation laws $\partial_t  q_a  + \partial_z  j_a  = 0$, where $q_a$ is density of the $a$'th conserved quantity $Q_a = \int \mathrm{d}z \: q_a$, while $j_a$ is its associated current.
Note, for the sake of generality, we will in the following let the subscript $a$ denote the index of the conserved charge.
Conventionally, the indices $a = 0, 1, 2$ correspond to particle number, momentum, and energy, respectively, such that $q_0 \equiv n$ and $q_2 \equiv \varepsilon$.
Their local thermodynamic expectation values are given in terms of the quasi-particle distribution via~\cite{PhysRevX.6.041065, PhysRevLett.117.207201}
\begin{equation}
\begin{aligned}
     q_a  &= \int_{-\infty}^{\infty} \mathrm{d}\theta \: \rho_{\mathrm{p}} (\theta) h_a(\theta) \; , \\
     j_a  &= \int_{-\infty}^{\infty} \mathrm{d}\theta \:  v^{\mathrm{eff}}(\theta) \rho_{\mathrm{p}} (\theta) h_a(\theta) \; .
\end{aligned}
\label{eq:conserved_charges}
\end{equation}
Here, $h_a$ is the single-particle eigenvalue of the $a$'th conserved quantity, which for atom number and energy read $h_0 = 1$ and $h_2 = \hbar^2\theta^2/2m$, respectively, while $v^{\mathrm{eff}}(\theta)$ is the local propagation velocity of a quasi-particle with rapidity $\theta$, given by
\begin{equation}
   v^{\mathrm{eff}}(\theta)= \frac{\hbar \theta}{m} + \int_{-\infty}^{\infty} d \theta^{\prime} \Delta\left(\theta-\theta^{\prime}\right)  \rho_{\mathrm{p}} (\theta') \left[v^{\mathrm{eff}}\left(\theta^{\prime}\right)-v^{\mathrm{eff}}(\theta)\right] \, .
\end{equation}
The propagation velocity is modified by interactions with other particles, which manifests as the Wigner delay time~\cite{PhysRev.98.145} associated with the quantum mechanical phase shifts occurring upon elastic collisions of the atoms.

The infinite number of continuity equations of integrable systems makes a conventional hydrodynamics treatment intractable; GHD overcomes this by expressing the dynamics via a collisionless Boltzmann equation for the quasi-particles~\cite{PhysRevX.6.041065, PhysRevLett.117.207201} 
\begin{equation}
    \partial_t \rho_{\mathrm{p}} + \partial_z (v^{\mathrm{eff}} \rho_{\mathrm{p}}) - \hbar^{-1} \partial_\theta ( \partial_z V \: \rho_{\mathrm{p}}) = 0 \; .
\label{eq:GHD_main}
\end{equation}
Eq.~\eqref{eq:GHD_main} facilitates numerical simulation of the large-scale dynamics of the system; all GHD simulations performed for this work solve Eq.~\eqref{eq:GHD_main} using backwards semi-Lagrangian schemes~\cite{MOLLER2023112431}, implemented in the iFluid library~\cite{10.21468/SciPostPhys.8.3.041}.  
Further, when comparing GHD simulations to experimental measurements, we account for the additional density dynamics during the 2~ms free expansion time.

\begin{figure*}
    \centering
    \includegraphics{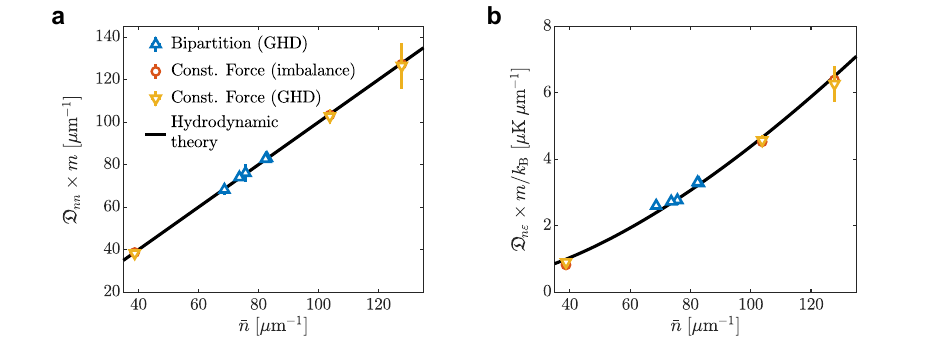}

    \phantomsubfloat{\label{fig:Drude_weights_GHD_a}}
    \phantomsubfloat{\label{fig:Drude_weights_GHD_b}}
    \vspace{-2\baselineskip}

    \caption{\label{fig:Drude_weights_GHD}
    \textbf{Drude Weights extracted from GHD simulations.}
    \textbf{a,b,} Drude weights characterising the linear response of the atomic current \textbf{(a)} and energy current \textbf{(b)} to a field coupled to the atomic density.
    The coefficients are extracted from GHD simulations of the experiment using Eqs.~\eqref{eq:Drudeweight_constforce} and~\eqref{eq:Drudeweight_bipartition}: For symbols labeled \textit{GHD}, the currents are calculated directly from the quasi-particle distribution using Eq.~\eqref{eq:conserved_charges}, while for those labeled \textit{imbalance}, the current is calculated via $\Delta N$.
    The former account for the full energy density, whereas the latter only consider the interaction energy density.
    The theoretical predictions for a thermal states at $T = 50\:\mathrm{nK}$ are calculated from Eq.~\eqref{eq:Drudeweight_GHD} and plotted as a solid line. 
    All results are plotted as a function of the background atomic density $\bar{n}$.
    Errorbars are obtained from simulations including the standard error of $\bar n$ and $\delta \mu$.
    }
\end{figure*}

In Fig.~\ref{fig:Drude_weights_GHD} the results of simulating the experiment with GHD and extracting the Drude weights using Eqs.~\eqref{eq:Drudeweight_constforce} and~\eqref{eq:Drudeweight_bipartition} are shown.
Although the external field perturbation $\delta\mu$ is finite, we find that the results agree very well with the analytic prediction of Eq.~\eqref{eq:Drudeweight_GHD} based on quasi-particle propagation in a homogeneous system.

\section{Conductivity and Drude weights of integrable models} 

The conductivity matrix $\sigma_{ab}$ quantifies the linear response of the current $j_b$ to an external perturbation of dynamics coupled to the charge density $q_a$~\cite{mahan2013many}.
Given an external force applied to a one-dimensional many-body system with equilibrium Hamiltonian $\mathcal{H}_0$, such that $\mathcal{H}(t) = \mathcal{H}_0 - \phi_a (t) \int\mathrm{d}z \: z q_a$, where the second term represents a linear potential with time dependent strength $\phi_a (t)$, the response of the current $j_b$ is
\begin{equation}
    j_b(t) = \int_{-\infty}^{t} \mathrm{d}t' \: \sigma_{a b}(t - t') \phi_a(t') \; .
\end{equation}
Taking the inverse Fourier transform of the conductivity~\eqref{eq:conductivity}, we see that the Drude weight gives a constant contribution $\sigma_{a b}(t) = \mathfrak{D}_{a b} + \ldots$, while the contribution of the regular part, which we neglect for now, may be time dependent.
Thus, given a constant force $\phi_a(t) = \phi_a$, the ballistic response of the current $j_b$ is a linear growth in time
\begin{equation}
    j_b(t) = \mathfrak{D}_{a b} \phi_a t + \ldots
\end{equation}

Furthermore, the conductivity of a system is connected to spreading of correlations in response to a local change in the initial state.
Consider the connected two-point functions of local conserved densities
\begin{equation}
    S_{a b}(z, t):=\left\langle q_a(z, t) q_b(0,0)\right\rangle^c \; .
\end{equation}
The late-time hydrodynamic spreading of $S_{a b}$ under such a local perturbation is governed by separate ballistic and diffusive contributions following~\cite{De_Nardis_2022}
\begin{equation}
    \frac{1}{2} \int \mathrm{d} z \:  z^2\left(S_{a b}(z, t)+S_{a b}(z,-t)\right)=\mathfrak{D}_{a b} t^2+\mathfrak{L}_{a b} t + O(t) \; ,
    \label{eq:correlation_spreading}
\end{equation}
given finite Drude weights $\mathfrak{D}_{a b}$ and Onsager matrix $\mathfrak{L}_{a b}$.
The Onsager matrix is another fundamental transport coefficient, which accounts for diffusive contributions arising from thermal fluctuations of the thermodynamic state and is defined as the zero-frequency limit of the regular part of the conductivity $ \mathfrak{L}_{a b}= \lim _{\omega \rightarrow 0} \sigma_{a b}^{\mathrm{reg}}(\omega)$~\cite{De_Nardis_2022}.

The response function, and thus the Drude weight, of a finite temperature system is given in terms of the time-averaged current-current autocorrelation function via the Kubo formula~\cite{Kubo}, yielding
\begin{equation}
    \mathfrak{D}_{a b} = \lim_{t \to \infty}  \frac{\beta}{2 t }  \int_{-\infty}^{\infty} \mathrm{d} z \int_{0}^{t} \mathrm{d} t' \left\langle j_a (z, t') j_b(0,0) \right\rangle^c \; .
    \label{eq:Kubo}
\end{equation}
Following GHD, analytic expressions for transport coefficients can be derived using the method of hydrodynamics projections, which in essence describes the emergent dynamics by projecting onto the subspace of slowly varying conserved densities and their gradients, enables evaluation of the current-current autocorrelation function of Eq.~\eqref{eq:Kubo}.
Thus, the Drude weights can be expressed in terms of the thermal distribution of quasi-particles via~\cite{10.21468/SciPostPhys.3.6.039}
\begin{equation}
    \mathfrak{D}_{a b} = \beta \int_{-\infty}^{\infty} \mathrm{d}\theta \: \rho_\mathrm{p}(\theta) \left( 1 - \vartheta(\theta) \right) \left( v^\mathrm{eff} (\theta )\right)^2 h_a^\mathrm{dr} (\theta) h_b^\mathrm{dr} (\theta) \; ,
    \label{eq:Drudeweight_GHD}
\end{equation}
where the superscript 'dr' denotes the dressing of a function, defined as
\begin{equation}
    h_a^{\mathrm{dr}}(\theta) = h_a(\theta)-\frac{1}{2 \pi} \int_{-\infty}^{\infty} \mathrm{d} \theta^{\prime} \Delta\left(\theta-\theta^{\prime}\right) \vartheta\left(\theta^{\prime}\right) h_a^{\mathrm{dr}}\left(\theta^{\prime}\right) .
\end{equation}
Like the Drude weight, an analytic expression for the Onsager matrix can be obtained via form factor expansion over the elementary excitations of the integrable model yielding~\cite{PhysRevLett.121.160603}
\begin{equation}
\begin{aligned}
\mathfrak{L}_{a b}=& \int \frac{\mathrm{d} \theta_1 \mathrm{~d} \theta_2}{2} \rho_{\mathrm{p}}\left(\theta_1\right) \left(1 - \vartheta(\theta_1)\right) \rho_{\mathrm{p} }\left(\theta_2\right) \left(1 - \vartheta(\theta_2)\right) \\
& \times \left|v^{\text {eff}}\left(\theta_1\right)-v^{\text {eff}}\left(\theta_2\right)\right| \left( \frac{\Delta^{\mathrm{dr}}\left(\theta_1, \theta_2\right)}{2 \pi}\right)^2 \\
& \times \bigg(\frac{h_{a}^{\mathrm{dr}}\left(\theta_2\right)}{\rho_{\mathrm{s}}\left(\theta_2\right)}-\frac{h_{a}^{\mathrm{dr}}\left(\theta_1\right)}{\rho_{\mathrm{s}}\left(\theta_1\right)}\bigg)\bigg(\frac{h_{b}^{\mathrm{dr}}\left(\theta_2\right)}{ \rho_{\mathrm{s}}\left(\theta_2\right)}-\frac{h_{b}^{\mathrm{dr}}\left(\theta_1\right)}{\rho_{\mathrm{s} }\left(\theta_1\right)}\bigg) \; .
\end{aligned}
\end{equation}
In integrable systems, diffusive contributions are generally subleading to ballistic, particularly in the quasi-condensate regime of the 1D Bose gas~\cite{10.21468/SciPostPhysCore.7.2.025}.
Knowing the large-scale spreading of correlations via Eq.~\eqref{eq:correlation_spreading}, we can estimate a time until which diffusion is dominant over ballistic spreading as $t_{a b}^\star = \mathfrak{L}_{a b} / \mathfrak{D}_{a b}$.
For density-density correlations in a typical thermal state of the experiment $t^\star \sim 10^{-40} \: \mathrm{s}$, whereby diffusion can be fully neglected.

\bibliography{references}

\begin{thebibliography}{50}%
\makeatletter
\providecommand \@ifxundefined [1]{%
 \@ifx{#1\undefined}
}%
\providecommand \@ifnum [1]{%
 \ifnum #1\expandafter \@firstoftwo
 \else \expandafter \@secondoftwo
 \fi
}%
\providecommand \@ifx [1]{%
 \ifx #1\expandafter \@firstoftwo
 \else \expandafter \@secondoftwo
 \fi
}%
\providecommand \natexlab [1]{#1}%
\providecommand \enquote  [1]{``#1''}%
\providecommand \bibnamefont  [1]{#1}%
\providecommand \bibfnamefont [1]{#1}%
\providecommand \citenamefont [1]{#1}%
\providecommand \href@noop [0]{\@secondoftwo}%
\providecommand \href [0]{\begingroup \@sanitize@url \@href}%
\providecommand \@href[1]{\@@startlink{#1}\@@href}%
\providecommand \@@href[1]{\endgroup#1\@@endlink}%
\providecommand \@sanitize@url [0]{\catcode `\\12\catcode `\$12\catcode `\&12\catcode `\#12\catcode `\^12\catcode `\_12\catcode `\%12\relax}%
\providecommand \@@startlink[1]{}%
\providecommand \@@endlink[0]{}%
\providecommand \url  [0]{\begingroup\@sanitize@url \@url }%
\providecommand \@url [1]{\endgroup\@href {#1}{\urlprefix }}%
\providecommand \urlprefix  [0]{URL }%
\providecommand \Eprint [0]{\href }%
\providecommand \doibase [0]{https://doi.org/}%
\providecommand \selectlanguage [0]{\@gobble}%
\providecommand \bibinfo  [0]{\@secondoftwo}%
\providecommand \bibfield  [0]{\@secondoftwo}%
\providecommand \translation [1]{[#1]}%
\providecommand \BibitemOpen [0]{}%
\providecommand \bibitemStop [0]{}%
\providecommand \bibitemNoStop [0]{.\EOS\space}%
\providecommand \EOS [0]{\spacefactor3000\relax}%
\providecommand \BibitemShut  [1]{\csname bibitem#1\endcsname}%
\let\auto@bib@innerbib\@empty
\bibitem [{\citenamefont {Scalapino}\ \emph {et~al.}(1993)\citenamefont {Scalapino}, \citenamefont {White},\ and\ \citenamefont {Zhang}}]{PhysRevB.47.7995}%
  \BibitemOpen
  \bibfield  {author} {\bibinfo {author} {\bibfnamefont {D.~J.}\ \bibnamefont {Scalapino}}, \bibinfo {author} {\bibfnamefont {S.~R.}\ \bibnamefont {White}},\ and\ \bibinfo {author} {\bibfnamefont {S.}~\bibnamefont {Zhang}},\ }\bibfield  {title} {\bibinfo {title} {Insulator, metal, or superconductor: The criteria},\ }\href {https://doi.org/10.1103/PhysRevB.47.7995} {\bibfield  {journal} {\bibinfo  {journal} {Phys. Rev. B}\ }\textbf {\bibinfo {volume} {47}},\ \bibinfo {pages} {7995} (\bibinfo {year} {1993})}\BibitemShut {NoStop}%
\bibitem [{\citenamefont {Drude}(1900)}]{Drude_Metalle}%
  \BibitemOpen
  \bibfield  {author} {\bibinfo {author} {\bibfnamefont {P.}~\bibnamefont {Drude}},\ }\bibfield  {title} {\bibinfo {title} {Zur {E}lektronentheorie der {M}etalle},\ }\href {https://doi.org/https://doi.org/10.1002/andp.19003060312} {\bibfield  {journal} {\bibinfo  {journal} {Annalen der Physik}\ }\textbf {\bibinfo {volume} {306}},\ \bibinfo {pages} {566} (\bibinfo {year} {1900})}\BibitemShut {NoStop}%
\bibitem [{\citenamefont {Kohn}(1964)}]{PhysRev.133.A171}%
  \BibitemOpen
  \bibfield  {author} {\bibinfo {author} {\bibfnamefont {W.}~\bibnamefont {Kohn}},\ }\bibfield  {title} {\bibinfo {title} {Theory of the insulating state},\ }\href {https://doi.org/10.1103/PhysRev.133.A171} {\bibfield  {journal} {\bibinfo  {journal} {Phys. Rev.}\ }\textbf {\bibinfo {volume} {133}},\ \bibinfo {pages} {A171} (\bibinfo {year} {1964})}\BibitemShut {NoStop}%
\bibitem [{\citenamefont {Horng}\ \emph {et~al.}(2011)\citenamefont {Horng}, \citenamefont {Chen}, \citenamefont {Geng}, \citenamefont {Girit}, \citenamefont {Zhang}, \citenamefont {Hao}, \citenamefont {Bechtel}, \citenamefont {Martin}, \citenamefont {Zettl}, \citenamefont {Crommie}, \citenamefont {Shen},\ and\ \citenamefont {Wang}}]{PhysRevB.83.165113}%
  \BibitemOpen
  \bibfield  {author} {\bibinfo {author} {\bibfnamefont {J.}~\bibnamefont {Horng}}, \bibinfo {author} {\bibfnamefont {C.-F.}\ \bibnamefont {Chen}}, \bibinfo {author} {\bibfnamefont {B.}~\bibnamefont {Geng}}, \bibinfo {author} {\bibfnamefont {C.}~\bibnamefont {Girit}}, \bibinfo {author} {\bibfnamefont {Y.}~\bibnamefont {Zhang}}, \bibinfo {author} {\bibfnamefont {Z.}~\bibnamefont {Hao}}, \bibinfo {author} {\bibfnamefont {H.~A.}\ \bibnamefont {Bechtel}}, \bibinfo {author} {\bibfnamefont {M.}~\bibnamefont {Martin}}, \bibinfo {author} {\bibfnamefont {A.}~\bibnamefont {Zettl}}, \bibinfo {author} {\bibfnamefont {M.~F.}\ \bibnamefont {Crommie}}, \bibinfo {author} {\bibfnamefont {Y.~R.}\ \bibnamefont {Shen}},\ and\ \bibinfo {author} {\bibfnamefont {F.}~\bibnamefont {Wang}},\ }\bibfield  {title} {\bibinfo {title} {Drude conductivity of {D}irac fermions in graphene},\ }\href {https://doi.org/10.1103/PhysRevB.83.165113} {\bibfield  {journal} {\bibinfo  {journal} {Phys. Rev. B}\ }\textbf {\bibinfo {volume} {83}},\
  \bibinfo {pages} {165113} (\bibinfo {year} {2011})}\BibitemShut {NoStop}%
\bibitem [{\citenamefont {Abedinpour}\ \emph {et~al.}(2011)\citenamefont {Abedinpour}, \citenamefont {Vignale}, \citenamefont {Principi}, \citenamefont {Polini}, \citenamefont {Tse},\ and\ \citenamefont {MacDonald}}]{PhysRevB.84.045429}%
  \BibitemOpen
  \bibfield  {author} {\bibinfo {author} {\bibfnamefont {S.~H.}\ \bibnamefont {Abedinpour}}, \bibinfo {author} {\bibfnamefont {G.}~\bibnamefont {Vignale}}, \bibinfo {author} {\bibfnamefont {A.}~\bibnamefont {Principi}}, \bibinfo {author} {\bibfnamefont {M.}~\bibnamefont {Polini}}, \bibinfo {author} {\bibfnamefont {W.-K.}\ \bibnamefont {Tse}},\ and\ \bibinfo {author} {\bibfnamefont {A.~H.}\ \bibnamefont {MacDonald}},\ }\bibfield  {title} {\bibinfo {title} {Drude weight, plasmon dispersion, and ac conductivity in doped graphene sheets},\ }\href {https://doi.org/10.1103/PhysRevB.84.045429} {\bibfield  {journal} {\bibinfo  {journal} {Phys. Rev. B}\ }\textbf {\bibinfo {volume} {84}},\ \bibinfo {pages} {045429} (\bibinfo {year} {2011})}\BibitemShut {NoStop}%
\bibitem [{\citenamefont {Frenzel}\ \emph {et~al.}(2014)\citenamefont {Frenzel}, \citenamefont {Lui}, \citenamefont {Shin}, \citenamefont {Kong},\ and\ \citenamefont {Gedik}}]{PhysRevLett.113.056602}%
  \BibitemOpen
  \bibfield  {author} {\bibinfo {author} {\bibfnamefont {A.~J.}\ \bibnamefont {Frenzel}}, \bibinfo {author} {\bibfnamefont {C.~H.}\ \bibnamefont {Lui}}, \bibinfo {author} {\bibfnamefont {Y.~C.}\ \bibnamefont {Shin}}, \bibinfo {author} {\bibfnamefont {J.}~\bibnamefont {Kong}},\ and\ \bibinfo {author} {\bibfnamefont {N.}~\bibnamefont {Gedik}},\ }\bibfield  {title} {\bibinfo {title} {Semiconducting-to-metallic photoconductivity crossover and temperature-dependent {D}rude weight in graphene},\ }\href {https://doi.org/10.1103/PhysRevLett.113.056602} {\bibfield  {journal} {\bibinfo  {journal} {Phys. Rev. Lett.}\ }\textbf {\bibinfo {volume} {113}},\ \bibinfo {pages} {056602} (\bibinfo {year} {2014})}\BibitemShut {NoStop}%
\bibitem [{\citenamefont {Zotos}\ and\ \citenamefont {Prelov\ifmmode~\check{s}\else \v{s}\fi{}ek}(1996)}]{PhysRevB.53.983}%
  \BibitemOpen
  \bibfield  {author} {\bibinfo {author} {\bibfnamefont {X.}~\bibnamefont {Zotos}}\ and\ \bibinfo {author} {\bibfnamefont {P.}~\bibnamefont {Prelov\ifmmode~\check{s}\else \v{s}\fi{}ek}},\ }\bibfield  {title} {\bibinfo {title} {Evidence for ideal insulating or conducting state in a one-dimensional integrable system},\ }\href {https://doi.org/10.1103/PhysRevB.53.983} {\bibfield  {journal} {\bibinfo  {journal} {Phys. Rev. B}\ }\textbf {\bibinfo {volume} {53}},\ \bibinfo {pages} {983} (\bibinfo {year} {1996})}\BibitemShut {NoStop}%
\bibitem [{\citenamefont {Castella}\ \emph {et~al.}(1995)\citenamefont {Castella}, \citenamefont {Zotos},\ and\ \citenamefont {Prelov\ifmmode~\check{s}\else \v{s}\fi{}ek}}]{PhysRevLett.74.972}%
  \BibitemOpen
  \bibfield  {author} {\bibinfo {author} {\bibfnamefont {H.}~\bibnamefont {Castella}}, \bibinfo {author} {\bibfnamefont {X.}~\bibnamefont {Zotos}},\ and\ \bibinfo {author} {\bibfnamefont {P.}~\bibnamefont {Prelov\ifmmode~\check{s}\else \v{s}\fi{}ek}},\ }\bibfield  {title} {\bibinfo {title} {Integrability and ideal conductance at finite temperatures},\ }\href {https://doi.org/10.1103/PhysRevLett.74.972} {\bibfield  {journal} {\bibinfo  {journal} {Phys. Rev. Lett.}\ }\textbf {\bibinfo {volume} {74}},\ \bibinfo {pages} {972} (\bibinfo {year} {1995})}\BibitemShut {NoStop}%
\bibitem [{\citenamefont {Castro-Alvaredo}\ \emph {et~al.}(2016)\citenamefont {Castro-Alvaredo}, \citenamefont {Doyon},\ and\ \citenamefont {Yoshimura}}]{PhysRevX.6.041065}%
  \BibitemOpen
  \bibfield  {author} {\bibinfo {author} {\bibfnamefont {O.~A.}\ \bibnamefont {Castro-Alvaredo}}, \bibinfo {author} {\bibfnamefont {B.}~\bibnamefont {Doyon}},\ and\ \bibinfo {author} {\bibfnamefont {T.}~\bibnamefont {Yoshimura}},\ }\bibfield  {title} {\bibinfo {title} {Emergent hydrodynamics in integrable quantum systems out of equilibrium},\ }\href {https://doi.org/10.1103/PhysRevX.6.041065} {\bibfield  {journal} {\bibinfo  {journal} {Phys. Rev. X}\ }\textbf {\bibinfo {volume} {6}},\ \bibinfo {pages} {041065} (\bibinfo {year} {2016})}\BibitemShut {NoStop}%
\bibitem [{\citenamefont {Bertini}\ \emph {et~al.}(2016)\citenamefont {Bertini}, \citenamefont {Collura}, \citenamefont {De~Nardis},\ and\ \citenamefont {Fagotti}}]{PhysRevLett.117.207201}%
  \BibitemOpen
  \bibfield  {author} {\bibinfo {author} {\bibfnamefont {B.}~\bibnamefont {Bertini}}, \bibinfo {author} {\bibfnamefont {M.}~\bibnamefont {Collura}}, \bibinfo {author} {\bibfnamefont {J.}~\bibnamefont {De~Nardis}},\ and\ \bibinfo {author} {\bibfnamefont {M.}~\bibnamefont {Fagotti}},\ }\bibfield  {title} {\bibinfo {title} {Transport in out-of-equilibrium {XXZ} chains: {E}xact profiles of charges and currents},\ }\href {https://doi.org/10.1103/PhysRevLett.117.207201} {\bibfield  {journal} {\bibinfo  {journal} {Phys. Rev. Lett.}\ }\textbf {\bibinfo {volume} {117}},\ \bibinfo {pages} {207201} (\bibinfo {year} {2016})}\BibitemShut {NoStop}%
\bibitem [{\citenamefont {Bulchandani}\ \emph {et~al.}(2018)\citenamefont {Bulchandani}, \citenamefont {Vasseur}, \citenamefont {Karrasch},\ and\ \citenamefont {Moore}}]{PhysRevB.97.045407}%
  \BibitemOpen
  \bibfield  {author} {\bibinfo {author} {\bibfnamefont {V.~B.}\ \bibnamefont {Bulchandani}}, \bibinfo {author} {\bibfnamefont {R.}~\bibnamefont {Vasseur}}, \bibinfo {author} {\bibfnamefont {C.}~\bibnamefont {Karrasch}},\ and\ \bibinfo {author} {\bibfnamefont {J.~E.}\ \bibnamefont {Moore}},\ }\bibfield  {title} {\bibinfo {title} {Bethe-{B}oltzmann hydrodynamics and spin transport in the {XXZ} chain},\ }\href {https://doi.org/10.1103/PhysRevB.97.045407} {\bibfield  {journal} {\bibinfo  {journal} {Phys. Rev. B}\ }\textbf {\bibinfo {volume} {97}},\ \bibinfo {pages} {045407} (\bibinfo {year} {2018})}\BibitemShut {NoStop}%
\bibitem [{\citenamefont {Nardis}\ \emph {et~al.}(2022)\citenamefont {Nardis}, \citenamefont {Doyon}, \citenamefont {Medenjak},\ and\ \citenamefont {Panfil}}]{De_Nardis_2022}%
  \BibitemOpen
  \bibfield  {author} {\bibinfo {author} {\bibfnamefont {J.~D.}\ \bibnamefont {Nardis}}, \bibinfo {author} {\bibfnamefont {B.}~\bibnamefont {Doyon}}, \bibinfo {author} {\bibfnamefont {M.}~\bibnamefont {Medenjak}},\ and\ \bibinfo {author} {\bibfnamefont {M.}~\bibnamefont {Panfil}},\ }\bibfield  {title} {\bibinfo {title} {Correlation functions and transport coefficients in generalised hydrodynamics},\ }\href {https://doi.org/10.1088/1742-5468/ac3658} {\bibfield  {journal} {\bibinfo  {journal} {Journal of Statistical Mechanics: Theory and Experiment}\ }\textbf {\bibinfo {volume} {2022}},\ \bibinfo {pages} {014002} (\bibinfo {year} {2022})}\BibitemShut {NoStop}%
\bibitem [{\citenamefont {Doyon}\ \emph {et~al.}(2023)\citenamefont {Doyon}, \citenamefont {Gopalakrishnan}, \citenamefont {M{\o}ller}, \citenamefont {Schmiedmayer},\ and\ \citenamefont {Vasseur}}]{doyon2023generalized}%
  \BibitemOpen
  \bibfield  {author} {\bibinfo {author} {\bibfnamefont {B.}~\bibnamefont {Doyon}}, \bibinfo {author} {\bibfnamefont {S.}~\bibnamefont {Gopalakrishnan}}, \bibinfo {author} {\bibfnamefont {F.}~\bibnamefont {M{\o}ller}}, \bibinfo {author} {\bibfnamefont {J.}~\bibnamefont {Schmiedmayer}},\ and\ \bibinfo {author} {\bibfnamefont {R.}~\bibnamefont {Vasseur}},\ }\href@noop {} {\bibinfo {title} {Generalized hydrodynamics: a perspective}} (\bibinfo {year} {2023}),\ \Eprint {https://arxiv.org/abs/2311.03438} {arXiv:2311.03438 [cond-mat.stat-mech]} \BibitemShut {NoStop}%
\bibitem [{\citenamefont {De~Nardis}\ \emph {et~al.}(2018)\citenamefont {De~Nardis}, \citenamefont {Bernard},\ and\ \citenamefont {Doyon}}]{PhysRevLett.121.160603}%
  \BibitemOpen
  \bibfield  {author} {\bibinfo {author} {\bibfnamefont {J.}~\bibnamefont {De~Nardis}}, \bibinfo {author} {\bibfnamefont {D.}~\bibnamefont {Bernard}},\ and\ \bibinfo {author} {\bibfnamefont {B.}~\bibnamefont {Doyon}},\ }\bibfield  {title} {\bibinfo {title} {Hydrodynamic diffusion in integrable systems},\ }\href {https://doi.org/10.1103/PhysRevLett.121.160603} {\bibfield  {journal} {\bibinfo  {journal} {Phys. Rev. Lett.}\ }\textbf {\bibinfo {volume} {121}},\ \bibinfo {pages} {160603} (\bibinfo {year} {2018})}\BibitemShut {NoStop}%
\bibitem [{\citenamefont {Gopalakrishnan}\ \emph {et~al.}(2018)\citenamefont {Gopalakrishnan}, \citenamefont {Huse}, \citenamefont {Khemani},\ and\ \citenamefont {Vasseur}}]{PhysRevB.98.220303}%
  \BibitemOpen
  \bibfield  {author} {\bibinfo {author} {\bibfnamefont {S.}~\bibnamefont {Gopalakrishnan}}, \bibinfo {author} {\bibfnamefont {D.~A.}\ \bibnamefont {Huse}}, \bibinfo {author} {\bibfnamefont {V.}~\bibnamefont {Khemani}},\ and\ \bibinfo {author} {\bibfnamefont {R.}~\bibnamefont {Vasseur}},\ }\bibfield  {title} {\bibinfo {title} {Hydrodynamics of operator spreading and quasiparticle diffusion in interacting integrable systems},\ }\href {https://doi.org/10.1103/PhysRevB.98.220303} {\bibfield  {journal} {\bibinfo  {journal} {Phys. Rev. B}\ }\textbf {\bibinfo {volume} {98}},\ \bibinfo {pages} {220303(R} (\bibinfo {year} {2018})}\BibitemShut {NoStop}%
\bibitem [{\citenamefont {Schemmer}\ \emph {et~al.}(2019)\citenamefont {Schemmer}, \citenamefont {Bouchoule}, \citenamefont {Doyon},\ and\ \citenamefont {Dubail}}]{schemmer2019generalized}%
  \BibitemOpen
  \bibfield  {author} {\bibinfo {author} {\bibfnamefont {M.}~\bibnamefont {Schemmer}}, \bibinfo {author} {\bibfnamefont {I.}~\bibnamefont {Bouchoule}}, \bibinfo {author} {\bibfnamefont {B.}~\bibnamefont {Doyon}},\ and\ \bibinfo {author} {\bibfnamefont {J.}~\bibnamefont {Dubail}},\ }\bibfield  {title} {\bibinfo {title} {Generalized {H}ydrodynamics on an atom chip},\ }\href {https://doi.org/10.1103/PhysRevLett.122.090601} {\bibfield  {journal} {\bibinfo  {journal} {Phys. Rev. Lett.}\ }\textbf {\bibinfo {volume} {122}},\ \bibinfo {pages} {090601} (\bibinfo {year} {2019})}\BibitemShut {NoStop}%
\bibitem [{\citenamefont {Malvania}\ \emph {et~al.}(2021)\citenamefont {Malvania}, \citenamefont {Zhang}, \citenamefont {Le}, \citenamefont {Dubail}, \citenamefont {Rigol},\ and\ \citenamefont {Weiss}}]{malvania2020generalized}%
  \BibitemOpen
  \bibfield  {author} {\bibinfo {author} {\bibfnamefont {N.}~\bibnamefont {Malvania}}, \bibinfo {author} {\bibfnamefont {Y.}~\bibnamefont {Zhang}}, \bibinfo {author} {\bibfnamefont {Y.}~\bibnamefont {Le}}, \bibinfo {author} {\bibfnamefont {J.}~\bibnamefont {Dubail}}, \bibinfo {author} {\bibfnamefont {M.}~\bibnamefont {Rigol}},\ and\ \bibinfo {author} {\bibfnamefont {D.~S.}\ \bibnamefont {Weiss}},\ }\bibfield  {title} {\bibinfo {title} {{G}eneralized {H}ydrodynamics in strongly interacting {1D} {B}ose gases},\ }\href {https://doi.org/10.1126/science.abf0147} {\bibfield  {journal} {\bibinfo  {journal} {Science}\ }\textbf {\bibinfo {volume} {373}},\ \bibinfo {pages} {1129} (\bibinfo {year} {2021})}\BibitemShut {NoStop}%
\bibitem [{\citenamefont {M\o{}ller}\ \emph {et~al.}(2021)\citenamefont {M\o{}ller}, \citenamefont {Li}, \citenamefont {Mazets}, \citenamefont {Stimming}, \citenamefont {Zhou}, \citenamefont {Zhu}, \citenamefont {Chen},\ and\ \citenamefont {Schmiedmayer}}]{PhysRevLett.126.090602}%
  \BibitemOpen
  \bibfield  {author} {\bibinfo {author} {\bibfnamefont {F.}~\bibnamefont {M\o{}ller}}, \bibinfo {author} {\bibfnamefont {C.}~\bibnamefont {Li}}, \bibinfo {author} {\bibfnamefont {I.}~\bibnamefont {Mazets}}, \bibinfo {author} {\bibfnamefont {H.-P.}\ \bibnamefont {Stimming}}, \bibinfo {author} {\bibfnamefont {T.}~\bibnamefont {Zhou}}, \bibinfo {author} {\bibfnamefont {Z.}~\bibnamefont {Zhu}}, \bibinfo {author} {\bibfnamefont {X.}~\bibnamefont {Chen}},\ and\ \bibinfo {author} {\bibfnamefont {J.}~\bibnamefont {Schmiedmayer}},\ }\bibfield  {title} {\bibinfo {title} {Extension of the generalized hydrodynamics to the dimensional crossover regime},\ }\href {https://doi.org/10.1103/PhysRevLett.126.090602} {\bibfield  {journal} {\bibinfo  {journal} {Phys. Rev. Lett.}\ }\textbf {\bibinfo {volume} {126}},\ \bibinfo {pages} {090602} (\bibinfo {year} {2021})}\BibitemShut {NoStop}%
\bibitem [{\citenamefont {Cataldini}\ \emph {et~al.}(2022)\citenamefont {Cataldini}, \citenamefont {M\o{}ller}, \citenamefont {Tajik}, \citenamefont {Sabino}, \citenamefont {Ji}, \citenamefont {Mazets}, \citenamefont {Schweigler}, \citenamefont {Rauer},\ and\ \citenamefont {Schmiedmayer}}]{PhysRevX.12.041032}%
  \BibitemOpen
  \bibfield  {author} {\bibinfo {author} {\bibfnamefont {F.}~\bibnamefont {Cataldini}}, \bibinfo {author} {\bibfnamefont {F.}~\bibnamefont {M\o{}ller}}, \bibinfo {author} {\bibfnamefont {M.}~\bibnamefont {Tajik}}, \bibinfo {author} {\bibfnamefont {J.~a.}\ \bibnamefont {Sabino}}, \bibinfo {author} {\bibfnamefont {S.-C.}\ \bibnamefont {Ji}}, \bibinfo {author} {\bibfnamefont {I.}~\bibnamefont {Mazets}}, \bibinfo {author} {\bibfnamefont {T.}~\bibnamefont {Schweigler}}, \bibinfo {author} {\bibfnamefont {B.}~\bibnamefont {Rauer}},\ and\ \bibinfo {author} {\bibfnamefont {J.}~\bibnamefont {Schmiedmayer}},\ }\bibfield  {title} {\bibinfo {title} {Emergent {P}auli blocking in a weakly interacting {B}ose gas},\ }\href {https://doi.org/10.1103/PhysRevX.12.041032} {\bibfield  {journal} {\bibinfo  {journal} {Phys. Rev. X}\ }\textbf {\bibinfo {volume} {12}},\ \bibinfo {pages} {041032} (\bibinfo {year} {2022})}\BibitemShut {NoStop}%
\bibitem [{\citenamefont {Reichel}\ and\ \citenamefont {Vuletic}(2011)}]{reichel2011atom}%
  \BibitemOpen
  \bibfield  {author} {\bibinfo {author} {\bibfnamefont {J.}~\bibnamefont {Reichel}}\ and\ \bibinfo {author} {\bibfnamefont {V.}~\bibnamefont {Vuletic}},\ }\href {https://books.google.at/books?id=IpBfSI7sS20C} {\emph {\bibinfo {title} {Atom Chips}}}\ (\bibinfo  {publisher} {Wiley},\ \bibinfo {year} {2011})\BibitemShut {NoStop}%
\bibitem [{\citenamefont {Lieb}\ and\ \citenamefont {Liniger}(1963)}]{lieb1963exact}%
  \BibitemOpen
  \bibfield  {author} {\bibinfo {author} {\bibfnamefont {E.~H.}\ \bibnamefont {Lieb}}\ and\ \bibinfo {author} {\bibfnamefont {W.}~\bibnamefont {Liniger}},\ }\bibfield  {title} {\bibinfo {title} {Exact analysis of an interacting {B}ose gas. {I}. {T}he general solution and the ground state},\ }\href {https://doi.org/10.1103/PhysRev.130.1605} {\bibfield  {journal} {\bibinfo  {journal} {Phys. Rev.}\ }\textbf {\bibinfo {volume} {130}},\ \bibinfo {pages} {1605} (\bibinfo {year} {1963})}\BibitemShut {NoStop}%
\bibitem [{\citenamefont {Olshanii}(1998)}]{PhysRevLett.81.938}%
  \BibitemOpen
  \bibfield  {author} {\bibinfo {author} {\bibfnamefont {M.}~\bibnamefont {Olshanii}},\ }\bibfield  {title} {\bibinfo {title} {Atomic scattering in the presence of an external confinement and a gas of impenetrable bosons},\ }\href {https://doi.org/10.1103/PhysRevLett.81.938} {\bibfield  {journal} {\bibinfo  {journal} {Phys. Rev. Lett.}\ }\textbf {\bibinfo {volume} {81}},\ \bibinfo {pages} {938} (\bibinfo {year} {1998})}\BibitemShut {NoStop}%
\bibitem [{\citenamefont {Kheruntsyan}\ \emph {et~al.}(2003)\citenamefont {Kheruntsyan}, \citenamefont {Gangardt}, \citenamefont {Drummond},\ and\ \citenamefont {Shlyapnikov}}]{PhysRevLett.91.040403}%
  \BibitemOpen
  \bibfield  {author} {\bibinfo {author} {\bibfnamefont {K.~V.}\ \bibnamefont {Kheruntsyan}}, \bibinfo {author} {\bibfnamefont {D.~M.}\ \bibnamefont {Gangardt}}, \bibinfo {author} {\bibfnamefont {P.~D.}\ \bibnamefont {Drummond}},\ and\ \bibinfo {author} {\bibfnamefont {G.~V.}\ \bibnamefont {Shlyapnikov}},\ }\bibfield  {title} {\bibinfo {title} {Pair correlations in a finite-temperature 1d {B}ose gas},\ }\href {https://doi.org/10.1103/PhysRevLett.91.040403} {\bibfield  {journal} {\bibinfo  {journal} {Phys. Rev. Lett.}\ }\textbf {\bibinfo {volume} {91}},\ \bibinfo {pages} {040403} (\bibinfo {year} {2003})}\BibitemShut {NoStop}%
\bibitem [{\citenamefont {Tajik}\ \emph {et~al.}(2019)\citenamefont {Tajik}, \citenamefont {Rauer}, \citenamefont {Schweigler}, \citenamefont {Cataldini}, \citenamefont {{a}o Sabino}, \citenamefont {M{\o}ller}, \citenamefont {Ji}, \citenamefont {Mazets},\ and\ \citenamefont {Schmiedmayer}}]{Tajik:19}%
  \BibitemOpen
  \bibfield  {author} {\bibinfo {author} {\bibfnamefont {M.}~\bibnamefont {Tajik}}, \bibinfo {author} {\bibfnamefont {B.}~\bibnamefont {Rauer}}, \bibinfo {author} {\bibfnamefont {T.}~\bibnamefont {Schweigler}}, \bibinfo {author} {\bibfnamefont {F.}~\bibnamefont {Cataldini}}, \bibinfo {author} {\bibfnamefont {J.}~\bibnamefont {{a}o Sabino}}, \bibinfo {author} {\bibfnamefont {F.~S.}\ \bibnamefont {M{\o}ller}}, \bibinfo {author} {\bibfnamefont {S.-C.}\ \bibnamefont {Ji}}, \bibinfo {author} {\bibfnamefont {I.~E.}\ \bibnamefont {Mazets}},\ and\ \bibinfo {author} {\bibfnamefont {J.}~\bibnamefont {Schmiedmayer}},\ }\bibfield  {title} {\bibinfo {title} {Designing arbitrary one-dimensional potentials on an atom chip},\ }\href {https://doi.org/10.1364/OE.27.033474} {\bibfield  {journal} {\bibinfo  {journal} {Opt. Express}\ }\textbf {\bibinfo {volume} {27}},\ \bibinfo {pages} {33474} (\bibinfo {year} {2019})}\BibitemShut {NoStop}%
\bibitem [{\citenamefont {Mahan}(2013)}]{mahan2013many}%
  \BibitemOpen
  \bibfield  {author} {\bibinfo {author} {\bibfnamefont {G.}~\bibnamefont {Mahan}},\ }\href {https://books.google.at/books?id=TFDUBwAAQBAJ} {\emph {\bibinfo {title} {Many-Particle Physics}}},\ Physics of Solids and Liquids\ (\bibinfo  {publisher} {Springer US},\ \bibinfo {year} {2013})\BibitemShut {NoStop}%
\bibitem [{\citenamefont {Friedman}\ \emph {et~al.}(2020)\citenamefont {Friedman}, \citenamefont {Gopalakrishnan},\ and\ \citenamefont {Vasseur}}]{PhysRevB.101.180302}%
  \BibitemOpen
  \bibfield  {author} {\bibinfo {author} {\bibfnamefont {A.~J.}\ \bibnamefont {Friedman}}, \bibinfo {author} {\bibfnamefont {S.}~\bibnamefont {Gopalakrishnan}},\ and\ \bibinfo {author} {\bibfnamefont {R.}~\bibnamefont {Vasseur}},\ }\bibfield  {title} {\bibinfo {title} {Diffusive hydrodynamics from integrability breaking},\ }\href {https://doi.org/10.1103/PhysRevB.101.180302} {\bibfield  {journal} {\bibinfo  {journal} {Phys. Rev. B}\ }\textbf {\bibinfo {volume} {101}},\ \bibinfo {pages} {180302(R)} (\bibinfo {year} {2020})}\BibitemShut {NoStop}%
\bibitem [{\citenamefont {Møller}\ \emph {et~al.}(2024)\citenamefont {Møller}, \citenamefont {Cataldini},\ and\ \citenamefont {Schmiedmayer}}]{10.21468/SciPostPhysCore.7.2.025}%
  \BibitemOpen
  \bibfield  {author} {\bibinfo {author} {\bibfnamefont {F.}~\bibnamefont {Møller}}, \bibinfo {author} {\bibfnamefont {F.}~\bibnamefont {Cataldini}},\ and\ \bibinfo {author} {\bibfnamefont {J.}~\bibnamefont {Schmiedmayer}},\ }\bibfield  {title} {\bibinfo {title} {{Identifying diffusive length scales in one-dimensional {B}ose gases}},\ }\href {https://doi.org/10.21468/SciPostPhysCore.7.2.025} {\bibfield  {journal} {\bibinfo  {journal} {SciPost Phys. Core}\ }\textbf {\bibinfo {volume} {7}},\ \bibinfo {pages} {025} (\bibinfo {year} {2024})}\BibitemShut {NoStop}%
\bibitem [{\citenamefont {Kubo}(1957)}]{Kubo}%
  \BibitemOpen
  \bibfield  {author} {\bibinfo {author} {\bibfnamefont {R.}~\bibnamefont {Kubo}},\ }\bibfield  {title} {\bibinfo {title} {Statistical-mechanical theory of irreversible processes. {I}. {G}eneral theory and simple applications to magnetic and conduction problems},\ }\href {https://doi.org/10.1143/JPSJ.12.570} {\bibfield  {journal} {\bibinfo  {journal} {Journal of the Physical Society of Japan}\ }\textbf {\bibinfo {volume} {12}},\ \bibinfo {pages} {570} (\bibinfo {year} {1957})}\BibitemShut {NoStop}%
\bibitem [{\citenamefont {Ilievski}\ and\ \citenamefont {Prosen}(2013)}]{Ilievski2013}%
  \BibitemOpen
  \bibfield  {author} {\bibinfo {author} {\bibfnamefont {E.}~\bibnamefont {Ilievski}}\ and\ \bibinfo {author} {\bibfnamefont {T.}~\bibnamefont {Prosen}},\ }\bibfield  {title} {\bibinfo {title} {Thermodyamic bounds on {D}rude weights in terms of almost-conserved quantities},\ }\href {https://doi.org/10.1007/s00220-012-1599-4} {\bibfield  {journal} {\bibinfo  {journal} {Communications in Mathematical Physics}\ }\textbf {\bibinfo {volume} {318}},\ \bibinfo {pages} {809} (\bibinfo {year} {2013})}\BibitemShut {NoStop}%
\bibitem [{\citenamefont {Brantut}\ \emph {et~al.}(2012)\citenamefont {Brantut}, \citenamefont {Meineke}, \citenamefont {Stadler}, \citenamefont {Krinner},\ and\ \citenamefont {Esslinger}}]{doi:10.1126/science.1223175}%
  \BibitemOpen
  \bibfield  {author} {\bibinfo {author} {\bibfnamefont {J.-P.}\ \bibnamefont {Brantut}}, \bibinfo {author} {\bibfnamefont {J.}~\bibnamefont {Meineke}}, \bibinfo {author} {\bibfnamefont {D.}~\bibnamefont {Stadler}}, \bibinfo {author} {\bibfnamefont {S.}~\bibnamefont {Krinner}},\ and\ \bibinfo {author} {\bibfnamefont {T.}~\bibnamefont {Esslinger}},\ }\bibfield  {title} {\bibinfo {title} {Conduction of ultracold fermions through a mesoscopic channel},\ }\href {https://doi.org/10.1126/science.1223175} {\bibfield  {journal} {\bibinfo  {journal} {Science}\ }\textbf {\bibinfo {volume} {337}},\ \bibinfo {pages} {1069} (\bibinfo {year} {2012})}\BibitemShut {NoStop}%
\bibitem [{\citenamefont {Husmann}\ \emph {et~al.}(2015)\citenamefont {Husmann}, \citenamefont {Uchino}, \citenamefont {Krinner}, \citenamefont {Lebrat}, \citenamefont {Giamarchi}, \citenamefont {Esslinger},\ and\ \citenamefont {Brantut}}]{doi:10.1126/science.aac9584}%
  \BibitemOpen
  \bibfield  {author} {\bibinfo {author} {\bibfnamefont {D.}~\bibnamefont {Husmann}}, \bibinfo {author} {\bibfnamefont {S.}~\bibnamefont {Uchino}}, \bibinfo {author} {\bibfnamefont {S.}~\bibnamefont {Krinner}}, \bibinfo {author} {\bibfnamefont {M.}~\bibnamefont {Lebrat}}, \bibinfo {author} {\bibfnamefont {T.}~\bibnamefont {Giamarchi}}, \bibinfo {author} {\bibfnamefont {T.}~\bibnamefont {Esslinger}},\ and\ \bibinfo {author} {\bibfnamefont {J.-P.}\ \bibnamefont {Brantut}},\ }\bibfield  {title} {\bibinfo {title} {Connecting strongly correlated superfluids by a quantum point contact},\ }\href {https://doi.org/10.1126/science.aac9584} {\bibfield  {journal} {\bibinfo  {journal} {Science}\ }\textbf {\bibinfo {volume} {350}},\ \bibinfo {pages} {1498} (\bibinfo {year} {2015})}\BibitemShut {NoStop}%
\bibitem [{\citenamefont {Ilievski}\ and\ \citenamefont {De~Nardis}(2017)}]{PhysRevLett.119.020602}%
  \BibitemOpen
  \bibfield  {author} {\bibinfo {author} {\bibfnamefont {E.}~\bibnamefont {Ilievski}}\ and\ \bibinfo {author} {\bibfnamefont {J.}~\bibnamefont {De~Nardis}},\ }\bibfield  {title} {\bibinfo {title} {Microscopic origin of ideal conductivity in integrable quantum models},\ }\href {https://doi.org/10.1103/PhysRevLett.119.020602} {\bibfield  {journal} {\bibinfo  {journal} {Phys. Rev. Lett.}\ }\textbf {\bibinfo {volume} {119}},\ \bibinfo {pages} {020602} (\bibinfo {year} {2017})}\BibitemShut {NoStop}%
\bibitem [{\citenamefont {Raissi}\ \emph {et~al.}(2019)\citenamefont {Raissi}, \citenamefont {Perdikaris},\ and\ \citenamefont {Karniadakis}}]{RAISSI2019686}%
  \BibitemOpen
  \bibfield  {author} {\bibinfo {author} {\bibfnamefont {M.}~\bibnamefont {Raissi}}, \bibinfo {author} {\bibfnamefont {P.}~\bibnamefont {Perdikaris}},\ and\ \bibinfo {author} {\bibfnamefont {G.}~\bibnamefont {Karniadakis}},\ }\bibfield  {title} {\bibinfo {title} {Physics-informed neural networks: A deep learning framework for solving forward and inverse problems involving nonlinear partial differential equations},\ }\href {https://doi.org/https://doi.org/10.1016/j.jcp.2018.10.045} {\bibfield  {journal} {\bibinfo  {journal} {Journal of Computational Physics}\ }\textbf {\bibinfo {volume} {378}},\ \bibinfo {pages} {686} (\bibinfo {year} {2019})}\BibitemShut {NoStop}%
\bibitem [{\citenamefont {Møller}\ and\ \citenamefont {Schmiedmayer}(2020)}]{10.21468/SciPostPhys.8.3.041}%
  \BibitemOpen
  \bibfield  {author} {\bibinfo {author} {\bibfnamefont {F.~S.}\ \bibnamefont {Møller}}\ and\ \bibinfo {author} {\bibfnamefont {J.}~\bibnamefont {Schmiedmayer}},\ }\bibfield  {title} {\bibinfo {title} {{Introducing iFluid: a numerical framework for solving hydrodynamical equations in integrable models}},\ }\href {https://doi.org/10.21468/SciPostPhys.8.3.041} {\bibfield  {journal} {\bibinfo  {journal} {SciPost Phys.}\ }\textbf {\bibinfo {volume} {8}},\ \bibinfo {pages} {041} (\bibinfo {year} {2020})}\BibitemShut {NoStop}%
\bibitem [{\citenamefont {Møller}\ \emph {et~al.}(2023)\citenamefont {Møller}, \citenamefont {Besse}, \citenamefont {Mazets}, \citenamefont {Stimming},\ and\ \citenamefont {Mauser}}]{MOLLER2023112431}%
  \BibitemOpen
  \bibfield  {author} {\bibinfo {author} {\bibfnamefont {F.}~\bibnamefont {Møller}}, \bibinfo {author} {\bibfnamefont {N.}~\bibnamefont {Besse}}, \bibinfo {author} {\bibfnamefont {I.}~\bibnamefont {Mazets}}, \bibinfo {author} {\bibfnamefont {H.}~\bibnamefont {Stimming}},\ and\ \bibinfo {author} {\bibfnamefont {N.}~\bibnamefont {Mauser}},\ }\bibfield  {title} {\bibinfo {title} {The dissipative generalized hydrodynamic equations and their numerical solution},\ }\href {https://doi.org/https://doi.org/10.1016/j.jcp.2023.112431} {\bibfield  {journal} {\bibinfo  {journal} {Journal of Computational Physics}\ }\textbf {\bibinfo {volume} {493}},\ \bibinfo {pages} {112431} (\bibinfo {year} {2023})}\BibitemShut {NoStop}%
\bibitem [{\citenamefont {Doyon}\ and\ \citenamefont {Spohn}(2017)}]{10.21468/SciPostPhys.3.6.039}%
  \BibitemOpen
  \bibfield  {author} {\bibinfo {author} {\bibfnamefont {B.}~\bibnamefont {Doyon}}\ and\ \bibinfo {author} {\bibfnamefont {H.}~\bibnamefont {Spohn}},\ }\bibfield  {title} {\bibinfo {title} {{Drude Weight for the {L}ieb-{L}iniger {B}ose Gas}},\ }\href {https://doi.org/10.21468/SciPostPhys.3.6.039} {\bibfield  {journal} {\bibinfo  {journal} {SciPost Phys.}\ }\textbf {\bibinfo {volume} {3}},\ \bibinfo {pages} {039} (\bibinfo {year} {2017})}\BibitemShut {NoStop}%
\bibitem [{\citenamefont {Doyon}\ and\ \citenamefont {Yoshimura}(2017)}]{10.21468/SciPostPhys.2.2.014}%
  \BibitemOpen
  \bibfield  {author} {\bibinfo {author} {\bibfnamefont {B.}~\bibnamefont {Doyon}}\ and\ \bibinfo {author} {\bibfnamefont {T.}~\bibnamefont {Yoshimura}},\ }\bibfield  {title} {\bibinfo {title} {{A note on {G}eneralized {H}ydrodynamics: inhomogeneous fields and other concepts}},\ }\href {https://doi.org/10.21468/SciPostPhys.2.2.014} {\bibfield  {journal} {\bibinfo  {journal} {SciPost Phys.}\ }\textbf {\bibinfo {volume} {2}},\ \bibinfo {pages} {014} (\bibinfo {year} {2017})}\BibitemShut {NoStop}%
\bibitem [{\citenamefont {Salasnich}\ \emph {et~al.}(2002)\citenamefont {Salasnich}, \citenamefont {Parola},\ and\ \citenamefont {Reatto}}]{PhysRevA.65.043614}%
  \BibitemOpen
  \bibfield  {author} {\bibinfo {author} {\bibfnamefont {L.}~\bibnamefont {Salasnich}}, \bibinfo {author} {\bibfnamefont {A.}~\bibnamefont {Parola}},\ and\ \bibinfo {author} {\bibfnamefont {L.}~\bibnamefont {Reatto}},\ }\bibfield  {title} {\bibinfo {title} {Effective wave equations for the dynamics of cigar-shaped and disk-shaped {B}ose condensates},\ }\href {https://doi.org/10.1103/PhysRevA.65.043614} {\bibfield  {journal} {\bibinfo  {journal} {Phys. Rev. A}\ }\textbf {\bibinfo {volume} {65}},\ \bibinfo {pages} {043614} (\bibinfo {year} {2002})}\BibitemShut {NoStop}%
\bibitem [{\citenamefont {Bulchandani}\ \emph {et~al.}(2021)\citenamefont {Bulchandani}, \citenamefont {Gopalakrishnan},\ and\ \citenamefont {Ilievski}}]{Bulchandani_2021}%
  \BibitemOpen
  \bibfield  {author} {\bibinfo {author} {\bibfnamefont {V.~B.}\ \bibnamefont {Bulchandani}}, \bibinfo {author} {\bibfnamefont {S.}~\bibnamefont {Gopalakrishnan}},\ and\ \bibinfo {author} {\bibfnamefont {E.}~\bibnamefont {Ilievski}},\ }\bibfield  {title} {\bibinfo {title} {Superdiffusion in spin chains},\ }\href {https://doi.org/10.1088/1742-5468/ac12c7} {\bibfield  {journal} {\bibinfo  {journal} {Journal of Statistical Mechanics: Theory and Experiment}\ }\textbf {\bibinfo {volume} {2021}},\ \bibinfo {pages} {084001} (\bibinfo {year} {2021})}\BibitemShut {NoStop}%
\bibitem [{\citenamefont {Prosen}(2011)}]{PhysRevLett.106.217206}%
  \BibitemOpen
  \bibfield  {author} {\bibinfo {author} {\bibfnamefont {T.~c.~v.}\ \bibnamefont {Prosen}},\ }\bibfield  {title} {\bibinfo {title} {Open {XXZ} spin chain: Nonequilibrium steady state and a strict bound on ballistic transport},\ }\href {https://doi.org/10.1103/PhysRevLett.106.217206} {\bibfield  {journal} {\bibinfo  {journal} {Phys. Rev. Lett.}\ }\textbf {\bibinfo {volume} {106}},\ \bibinfo {pages} {217206} (\bibinfo {year} {2011})}\BibitemShut {NoStop}%
\bibitem [{\citenamefont {Nagy}\ \emph {et~al.}(2023)\citenamefont {Nagy}, \citenamefont {Kormos},\ and\ \citenamefont {Tak\'acs}}]{PhysRevB.108.L241105}%
  \BibitemOpen
  \bibfield  {author} {\bibinfo {author} {\bibfnamefont {B.~C.}\ \bibnamefont {Nagy}}, \bibinfo {author} {\bibfnamefont {M.}~\bibnamefont {Kormos}},\ and\ \bibinfo {author} {\bibfnamefont {G.}~\bibnamefont {Tak\'acs}},\ }\bibfield  {title} {\bibinfo {title} {Thermodynamics and fractal {D}rude weights in the sine-{G}ordon model},\ }\href {https://doi.org/10.1103/PhysRevB.108.L241105} {\bibfield  {journal} {\bibinfo  {journal} {Phys. Rev. B}\ }\textbf {\bibinfo {volume} {108}},\ \bibinfo {pages} {L241105} (\bibinfo {year} {2023})}\BibitemShut {NoStop}%
\bibitem [{\citenamefont {Rigol}\ \emph {et~al.}(2007)\citenamefont {Rigol}, \citenamefont {Dunjko}, \citenamefont {Yurovsky},\ and\ \citenamefont {Olshanii}}]{PhysRevLett.98.050405}%
  \BibitemOpen
  \bibfield  {author} {\bibinfo {author} {\bibfnamefont {M.}~\bibnamefont {Rigol}}, \bibinfo {author} {\bibfnamefont {V.}~\bibnamefont {Dunjko}}, \bibinfo {author} {\bibfnamefont {V.}~\bibnamefont {Yurovsky}},\ and\ \bibinfo {author} {\bibfnamefont {M.}~\bibnamefont {Olshanii}},\ }\bibfield  {title} {\bibinfo {title} {Relaxation in a completely integrable many-body quantum system: An ab initio study of the dynamics of the highly excited states of 1d lattice hard-core bosons},\ }\href {https://doi.org/10.1103/PhysRevLett.98.050405} {\bibfield  {journal} {\bibinfo  {journal} {Phys. Rev. Lett.}\ }\textbf {\bibinfo {volume} {98}},\ \bibinfo {pages} {050405} (\bibinfo {year} {2007})}\BibitemShut {NoStop}%
\bibitem [{\citenamefont {Langen}\ \emph {et~al.}(2015)\citenamefont {Langen}, \citenamefont {Erne}, \citenamefont {Geiger}, \citenamefont {Rauer}, \citenamefont {Schweigler}, \citenamefont {Kuhnert}, \citenamefont {Rohringer}, \citenamefont {Mazets}, \citenamefont {Gasenzer},\ and\ \citenamefont {Schmiedmayer}}]{doi:10.1126/science.1257026}%
  \BibitemOpen
  \bibfield  {author} {\bibinfo {author} {\bibfnamefont {T.}~\bibnamefont {Langen}}, \bibinfo {author} {\bibfnamefont {S.}~\bibnamefont {Erne}}, \bibinfo {author} {\bibfnamefont {R.}~\bibnamefont {Geiger}}, \bibinfo {author} {\bibfnamefont {B.}~\bibnamefont {Rauer}}, \bibinfo {author} {\bibfnamefont {T.}~\bibnamefont {Schweigler}}, \bibinfo {author} {\bibfnamefont {M.}~\bibnamefont {Kuhnert}}, \bibinfo {author} {\bibfnamefont {W.}~\bibnamefont {Rohringer}}, \bibinfo {author} {\bibfnamefont {I.~E.}\ \bibnamefont {Mazets}}, \bibinfo {author} {\bibfnamefont {T.}~\bibnamefont {Gasenzer}},\ and\ \bibinfo {author} {\bibfnamefont {J.}~\bibnamefont {Schmiedmayer}},\ }\bibfield  {title} {\bibinfo {title} {Experimental observation of a generalized {G}ibbs ensemble},\ }\href {https://doi.org/10.1126/science.1257026} {\bibfield  {journal} {\bibinfo  {journal} {Science}\ }\textbf {\bibinfo {volume} {348}},\ \bibinfo {pages} {207} (\bibinfo {year} {2015})}\BibitemShut {NoStop}%
\bibitem [{\citenamefont {Tang}\ \emph {et~al.}(2018)\citenamefont {Tang}, \citenamefont {Kao}, \citenamefont {Li}, \citenamefont {Seo}, \citenamefont {Mallayya}, \citenamefont {Rigol}, \citenamefont {Gopalakrishnan},\ and\ \citenamefont {Lev}}]{PhysRevX.8.021030}%
  \BibitemOpen
  \bibfield  {author} {\bibinfo {author} {\bibfnamefont {Y.}~\bibnamefont {Tang}}, \bibinfo {author} {\bibfnamefont {W.}~\bibnamefont {Kao}}, \bibinfo {author} {\bibfnamefont {K.-Y.}\ \bibnamefont {Li}}, \bibinfo {author} {\bibfnamefont {S.}~\bibnamefont {Seo}}, \bibinfo {author} {\bibfnamefont {K.}~\bibnamefont {Mallayya}}, \bibinfo {author} {\bibfnamefont {M.}~\bibnamefont {Rigol}}, \bibinfo {author} {\bibfnamefont {S.}~\bibnamefont {Gopalakrishnan}},\ and\ \bibinfo {author} {\bibfnamefont {B.~L.}\ \bibnamefont {Lev}},\ }\bibfield  {title} {\bibinfo {title} {Thermalization near integrability in a dipolar quantum {N}ewton's cradle},\ }\href {https://doi.org/10.1103/PhysRevX.8.021030} {\bibfield  {journal} {\bibinfo  {journal} {Phys. Rev. X}\ }\textbf {\bibinfo {volume} {8}},\ \bibinfo {pages} {021030} (\bibinfo {year} {2018})}\BibitemShut {NoStop}%
\bibitem [{\citenamefont {Bastianello}\ \emph {et~al.}(2021)\citenamefont {Bastianello}, \citenamefont {Luca},\ and\ \citenamefont {Vasseur}}]{Bastianello_2021}%
  \BibitemOpen
  \bibfield  {author} {\bibinfo {author} {\bibfnamefont {A.}~\bibnamefont {Bastianello}}, \bibinfo {author} {\bibfnamefont {A.~D.}\ \bibnamefont {Luca}},\ and\ \bibinfo {author} {\bibfnamefont {R.}~\bibnamefont {Vasseur}},\ }\bibfield  {title} {\bibinfo {title} {Hydrodynamics of weak integrability breaking},\ }\href {https://doi.org/10.1088/1742-5468/ac26b2} {\bibfield  {journal} {\bibinfo  {journal} {Journal of Statistical Mechanics: Theory and Experiment}\ }\textbf {\bibinfo {volume} {2021}},\ \bibinfo {pages} {114003} (\bibinfo {year} {2021})}\BibitemShut {NoStop}%
\bibitem [{\citenamefont {Manz}\ \emph {et~al.}(2010)\citenamefont {Manz}, \citenamefont {B\"ucker}, \citenamefont {Betz}, \citenamefont {Koller}, \citenamefont {Hofferberth}, \citenamefont {Mazets}, \citenamefont {Imambekov}, \citenamefont {Demler}, \citenamefont {Perrin}, \citenamefont {Schmiedmayer},\ and\ \citenamefont {Schumm}}]{PhysRevA.81.031610}%
  \BibitemOpen
  \bibfield  {author} {\bibinfo {author} {\bibfnamefont {S.}~\bibnamefont {Manz}}, \bibinfo {author} {\bibfnamefont {R.}~\bibnamefont {B\"ucker}}, \bibinfo {author} {\bibfnamefont {T.}~\bibnamefont {Betz}}, \bibinfo {author} {\bibfnamefont {C.}~\bibnamefont {Koller}}, \bibinfo {author} {\bibfnamefont {S.}~\bibnamefont {Hofferberth}}, \bibinfo {author} {\bibfnamefont {I.~E.}\ \bibnamefont {Mazets}}, \bibinfo {author} {\bibfnamefont {A.}~\bibnamefont {Imambekov}}, \bibinfo {author} {\bibfnamefont {E.}~\bibnamefont {Demler}}, \bibinfo {author} {\bibfnamefont {A.}~\bibnamefont {Perrin}}, \bibinfo {author} {\bibfnamefont {J.}~\bibnamefont {Schmiedmayer}},\ and\ \bibinfo {author} {\bibfnamefont {T.}~\bibnamefont {Schumm}},\ }\bibfield  {title} {\bibinfo {title} {Two-point density correlations of quasicondensates in free expansion},\ }\href {https://doi.org/10.1103/PhysRevA.81.031610} {\bibfield  {journal} {\bibinfo  {journal} {Phys. Rev. A}\ }\textbf {\bibinfo {volume} {81}},\ \bibinfo {pages} {031610(R)} (\bibinfo
  {year} {2010})}\BibitemShut {NoStop}%
\bibitem [{\citenamefont {Li}\ \emph {et~al.}(2023)\citenamefont {Li}, \citenamefont {Zhang}, \citenamefont {Yang}, \citenamefont {Lin}, \citenamefont {Gopalakrishnan}, \citenamefont {Rigol},\ and\ \citenamefont {Lev}}]{PhysRevA.107.L061302}%
  \BibitemOpen
  \bibfield  {author} {\bibinfo {author} {\bibfnamefont {K.-Y.}\ \bibnamefont {Li}}, \bibinfo {author} {\bibfnamefont {Y.}~\bibnamefont {Zhang}}, \bibinfo {author} {\bibfnamefont {K.}~\bibnamefont {Yang}}, \bibinfo {author} {\bibfnamefont {K.-Y.}\ \bibnamefont {Lin}}, \bibinfo {author} {\bibfnamefont {S.}~\bibnamefont {Gopalakrishnan}}, \bibinfo {author} {\bibfnamefont {M.}~\bibnamefont {Rigol}},\ and\ \bibinfo {author} {\bibfnamefont {B.~L.}\ \bibnamefont {Lev}},\ }\bibfield  {title} {\bibinfo {title} {Rapidity and momentum distributions of one-dimensional dipolar quantum gases},\ }\href {https://doi.org/10.1103/PhysRevA.107.L061302} {\bibfield  {journal} {\bibinfo  {journal} {Phys. Rev. A}\ }\textbf {\bibinfo {volume} {107}},\ \bibinfo {pages} {L061302} (\bibinfo {year} {2023})}\BibitemShut {NoStop}%
\bibitem [{\citenamefont {Baydin}\ \emph {et~al.}(2018)\citenamefont {Baydin}, \citenamefont {Pearlmutter}, \citenamefont {Radul},\ and\ \citenamefont {Siskind}}]{JMLR:v18:17-468}%
  \BibitemOpen
  \bibfield  {author} {\bibinfo {author} {\bibfnamefont {A.~G.}\ \bibnamefont {Baydin}}, \bibinfo {author} {\bibfnamefont {B.~A.}\ \bibnamefont {Pearlmutter}}, \bibinfo {author} {\bibfnamefont {A.~A.}\ \bibnamefont {Radul}},\ and\ \bibinfo {author} {\bibfnamefont {J.~M.}\ \bibnamefont {Siskind}},\ }\bibfield  {title} {\bibinfo {title} {Automatic differentiation in machine learning: a survey},\ }\href {http://jmlr.org/papers/v18/17-468.html} {\bibfield  {journal} {\bibinfo  {journal} {Journal of Machine Learning Research}\ }\textbf {\bibinfo {volume} {18}},\ \bibinfo {pages} {1} (\bibinfo {year} {2018})}\BibitemShut {NoStop}%
\bibitem [{\citenamefont {Yang}\ and\ \citenamefont {Yang}(1969)}]{10.1063/1.1664947}%
  \BibitemOpen
  \bibfield  {author} {\bibinfo {author} {\bibfnamefont {C.~N.}\ \bibnamefont {Yang}}\ and\ \bibinfo {author} {\bibfnamefont {C.~P.}\ \bibnamefont {Yang}},\ }\bibfield  {title} {\bibinfo {title} {{Thermodynamics of a One‐Dimensional System of Bosons with Repulsive Delta‐Function Interaction}},\ }\href {https://doi.org/10.1063/1.1664947} {\bibfield  {journal} {\bibinfo  {journal} {Journal of Mathematical Physics}\ }\textbf {\bibinfo {volume} {10}},\ \bibinfo {pages} {1115} (\bibinfo {year} {1969})}\BibitemShut {NoStop}%
\bibitem [{\citenamefont {Wigner}(1955)}]{PhysRev.98.145}%
  \BibitemOpen
  \bibfield  {author} {\bibinfo {author} {\bibfnamefont {E.~P.}\ \bibnamefont {Wigner}},\ }\bibfield  {title} {\bibinfo {title} {Lower limit for the energy derivative of the scattering phase shift},\ }\href {https://doi.org/10.1103/PhysRev.98.145} {\bibfield  {journal} {\bibinfo  {journal} {Phys. Rev.}\ }\textbf {\bibinfo {volume} {98}},\ \bibinfo {pages} {145} (\bibinfo {year} {1955})}\BibitemShut {NoStop}%
\end{thebibliography}%

\end{document}